\documentclass[journal=jpcafh,manuscript=article]{achemso}
\usepackage[utf8]{inputenc}
\usepackage{longtable}
\usepackage{multirow}
\usepackage{amsmath}
\usepackage{subfigure}
\usepackage[usenames,dvipsnames]{xcolor}
\usepackage{soul}
\usepackage{bm}

\usepackage{amssymb}
\usepackage{siunitx}
\usepackage{multicol} \usepackage{wrapfig} \usepackage{enumitem}
\usepackage{bm} \usepackage{outlines} 
\usepackage[normalem]{ulem}
\usepackage{url}
\usepackage{hyperref}
\usepackage{cancel}
\usepackage{comment}
\usepackage{siunitx}\usepackage{booktabs}

\SectionNumbersOn
\author{Sena Aydin} \affiliation[University of Basel]{Department of
  Chemistry, University of Basel, Klingelbergstrasse 80, CH-4056
  Basel, Switzerland.}

\author{Markus Meuwly} \affiliation[University of Basel]{Department of
  Chemistry, University of Basel, Klingelbergstrasse 80, CH-4056
  Basel, Switzerland.}  \email{m.meuwly@unibas.ch}

\title{Site-Specific Vibrational Dynamics to Probe Local and Global
  Protein Motions}

\makeatother
\begin{document}
\date{\today}\\

\begin{abstract}
Vibrational spectroscopic probes can provide site-specific information
on protein structure and dynamics. In this work, the possibility to
relate protein motion with the vibrational response for --SCN,
--N$_3$, and --SNO labels covalently linked to all alanine-residues in
lysozyme is investigated. Depending on the position of the probe along
the polypeptide chain, its local hydration, and the dynamics of its
environment, the vibrational response can encode not only local
fluctuations but also low-frequency, collective motions of the
protein. The spectroscopic probes are described using
machine-learning-based models for both bonded interactions,
represented by reproducing-kernel models, and electrostatic
interactions, represented by fluctuating minimally distributed
charges. Spectroscopic dynamics are characterized through frequency
fluctuation correlation functions (FFCFs). For many probe locations,
the FFCFs contain a non-decaying component on the simulation time
scale, indicating residual slow dynamics and incomplete sampling of
the underlying conformational fluctuations. The magnitude of these
static contributions is consistent with previous experimental
observations, providing qualitative validation of the
simulations. Overall, the results demonstrate that site-specific
vibrational probes can report on dynamics extending beyond their
immediate local environment and can, at suitable locations, provide
information on collective protein motions.
\end{abstract}

\maketitle

\section{Introduction}
Vibrational spectroscopy provides a complementary approach to probe
biomolecular dynamics, as it is sensitive to chemical
environments,\cite{cho:2009,middleton:2007} capturing both intra- and
intermolecular interactions,\cite{zheng:2007,zwier:1996} vibrational
couplings,\cite{hamm:1998} and conformational
changes.\cite{brewer:2007} In particular, the usage of site-specific
vibrational probes enables the selective observation of dynamics at
specified positions within a protein, providing detailed insight into
local structural changes and interactions which are valuable for
understanding the spatial organization and structural dynamics of
biological systems.\cite{kim:2013,ma:2015} The effectiveness of a
site-specific vibrational probe depends critically on its ability to
sensitively reflect its local chemical environment allowing the probe
to report subtle variations in local structure changes,
electrostatics, and hydrogen-bonding
interactions.\cite{Ramos:2019,taskent:2010,suydam:2003} Specific
efforts concern the relationship between molecular dynamics and
spectroscopic observables, studies of vibrational energy flow across
heme-protein and protein-water interfaces, and questions relating to
amide-I maps linking fluctuating vibrational frequencies to solvation,
electrostatics, conformational change, and protein structural
dynamics.\cite{straub:2011,straub:2014,straub:2015,straub:2020}\\

\noindent
The spectroscopic response of proteins in solution covers frequency
ranges up to $\nu \sim 1700$ cm$^{-1}$ and above $\sim 2800$
cm$^{-1}$. Therefore, vibrational labels ideally absorb in the window
$1700 \leq \nu \leq 2800$ cm$^{-1}$.\cite{gai:2011,hamm.rev:2015} A
number of such probes has been proposed in the past, including
cyanophenylalanine\cite{thielges:2015}, nitrile-derivatized amino
acids,\cite{gai:2003} the sulfhydryl band of
cysteines,\cite{hamm:2008} or deuterated carbons.\cite{romesberg:2011}
Also, non-natural labels consisting of metal-tricarbonyl modified with
a -(CH$_2$)$_{\rm n}$- linker,\cite{zanni:2013} nitrile
labels,\cite{fayer.ribo:2012} cyano\cite{romesberg.cn:2011} and
--SCN\cite{bredenbeck:2014} groups, or cyanamide\cite{cho:2018} were
considered. For the specific case of lysozyme, which is the protein
used in the present work, labeling with a ruthenium carbonyl complex
provided deeper understanding of the water dynamics from 2D-infrared
(2D-IR) experiments.\cite{kevin:2012,kevin.2:2012,kevin:2014}
Remarkably, it has been possible to demonstrate dynamic hydration and
effects of collective protein hydration extending over distances 20
\AA\/ away from the protein surface, consistent with recent
simulations on hydrated hemoglobin.\cite{MM.hb:2018,MM.hb:2020}\\

\noindent
The nitrile --CN stretching vibration is among the most widely
employed site-specific vibrational probes, as it absorbs in a
spectrally uncluttered region of the infrared spectrum and can be
readily incorporated into biological systems, including through
structurally related derivatives such as thiocyanate
(--SCN).\cite{nitrile.fang:2008,nitrile.li:2025,nitrile.lindquist:2009}
Nitrile-based probes retain sensitivity to local electrostatics,
hydrogen bonding, and solvation, thereby enabling detailed
characterization of protein environments and
dynamics.\cite{suydam:2003,fafarman:2006} In recent work, --SCN labels
attached to proteins residues (e.g., alanine in lysozyme) were shown
to exhibit distinct, site-dependent frequency shifts and sensitivity
to local hydration and structural changes, highlighting their utility
as probes of protein dynamics.\cite{MM.scn:2024} Similarly, the azide
stretching mode has been extensively utilized due to its strong and
well-isolated absorption feature and pronounced sensitivity to local
electrostatic
environments.\cite{MM.n3:2022,azide.bazewicz:2013,MM.n3:2021}\\

\noindent
Another useful spectroscopic label is the azide group, --N$_3$. The
noncanonical amino acid AHA absorbs around $\sim 2100$ cm$^{-1}$ with
a comparatively large extinction coefficient of up to 400
M$^{-1}$cm$^{-1}$.\cite{hamm:2012} From a preparative perspective
attachment of --N$_3$ to alanine (to give AlaN$_3$) and AHA and
incorporation at almost any position of a protein through known
expression techniques has been demonstrated.\cite{bertozzi:2002}
Furthermore, attachment of an --N$_3$ probe is a spatially small
modification and the chemical perturbations induced are expected to be
small. This makes AlaN$_3$ and AHA worthwhile modifications to probe
local protein dynamics. For AHA it has been demonstrated that it can
be used to characterize the recognition site between the PDZ2 domain
and its binding partner to provide site-specific insight into the
underlying mechanisms of how signaling proteins
function.\cite{stock:2018}\\

\noindent
Finally, S‑nitrosothiol (--SNO) functionalities, formed by the
covalent attachment of nitric oxide (NO) to a cysteine thiol,
represent a biologically significant post‑translational modification
with wide‑ranging regulatory roles in cellular signaling and protein
function. Protein S‑nitrosylation has been recognized as a major redox
signaling mechanism, with reversible SNO formation influencing
activity, stability and interactions of target
proteins.\cite{WILLIAMS:2003} The --SNO group has distinct vibrational
features, including the S–N stretch, NO stretch, and SNO bending
modes, in S‑nitrosylated myoglobin, demonstrating that these modes are
spectroscopically resolvable and sensitive to local structural and
hydration environments.\cite{MM.sno:2021} Likewise, gas‑phase infrared
multiple photon dissociation spectroscopy of protonated
S‑nitrosoglutathione has established clear vibrational signatures
associated with the --SNO moiety, providing benchmarks for probing
S‑nitrosothiol structure and dynamics.\cite{gregori:2024}\\

\noindent
The present study aims at a direct comparison of three different
vibrational probes (--SCN, --N$_3$, and --SNO) to characterize
site-specific dynamics in lysozyme. A particular focus is on the
question whether and to what degree the high-frequency modes of the
three labels report on the low-frequency protein dynamics. For this,
dedicated models capturing geometry-dependent electrostatics for the
three labels are developed and optimized to best capture
intermolecular interactions with water.\\

\noindent
The work is structured as follows. First, the methods are described,
followed by the results of the force field adjustments and an analysis
of the frequency correlation functions. Next, the spectroscopy of the
labels in the context of protein dynamics is discussed. This combined
approach provides a detailed, atomically resolved characterization of
how distinct microenvironments within lysozyme influence probe
responses and dynamics. Finally, conclusions are drawn.\\

\section{Computational Methods}

\subsection{Potential Energy Surface}
For the majority of the simulation system, the
CGenFF\cite{cgenff:2012} force field as implemented in the CHARMM
program suite\cite{charmm:2009,MM.charmm:2024}, was employed together
with the TIP3P water\cite{tip3p_1983} model to ensure consistency. In
contrast, the spectroscopic label was treated with a significantly
more advanced energy description. Previously developed reproducing
kernel-based potential energy surfaces (RKHS) models
\cite{rabitz:1996,MM.rkhs:2017} for --SCN and --N$_3$ probes were
employed in earlier studies.\cite{MM.scn:2024,MM.n3:2021} In contrast,
in the present work a new PES model has been constructed specifically
for the –SNO label. Starting from an MP2/aug-cc-pVTZ-optimized
structure of S-nitrosocysteine a detailed potential energy surface
(PES) for the –SNO group was computed at the pair natural
orbital-based coupled-cluster level, PNO-LCCSD(T)-F12
\cite{lccsd-schwilk-2017,lccsdf12-schwilk-2017} using the aug-cc-pVTZ
basis set as implemented in MOLPRO \cite{molpro:2011}.\\

\noindent
The \textit{ab initio} PES was generated on a grid defined in Jacobi
coordinates ($R$, $r$, and $\theta$). Here, $r$ denotes the N-O
distance, $R$ the distance between the sulfur atom and the center mass
of the N-O fragment and $\theta$ was sampled at 7 points between
-20$^{\circ}$ and 20$^{\circ}$, while the radial coordinates were
sampled at 16 points for $r$ between --0.25 and 0.5 \AA\/ around the
minimum energy structure and 13 points for $R$ spanning -0.5 and 0.5
\AA\/ around the equilibrium structure. These coordinates have been
previously been shown to be advantageous for spectroscopic
modelling. Figure \ref{sifig:fig1} illustrates the RKHS representation
of the PES constructed from 1273 of the 1456 ab initio energies at
$\theta = 129.7^{\circ}$ and the distribution of potential energy
error.\\

\subsection{Electrostatic Models}
Flexible MDCM (fMDCM)\cite{MM.fmdcm:2022} is a charge model that
allows for the relocation of charges within a minimally distributed
charge model (MDCM) with respect to their reference geometry. This
method captures changes in the molecular charge distribution as a
function of geometry. The fMDCM approach is validated on a number of
small molecules. It provides accuracies in the electrostatic potential
(ESP) of 0.5 kcal/mol on average compared with reference data from
electronic structure calculations. In contrast,
MDCM\cite{mdcm.devereux:2020} and point charges have higher root mean
squared errors.\cite{MM.fmdcm:2022,MM.ff:2025}\\

\noindent
Here, individual fMDCM models were generated for the --SCN, --N$_3$,
and --SNO labels. These models were developed based on the CH$_3$–X
molecule (where X represents the label). The total charge of each
system was zero. To obtain the models, an initial MDCM model was
generated, which was then used as a reference to create the
corresponding fMDCM models for each label. Initial MDCM models
consisting of 10, 11, and 8 charges were generated for the --SCN,
--SNO, and --N$_3$ labels, respectively. These models served as
references for the subsequent fMDCM fitting procedure, resulting in
final fMDCM models containing 9, 10, and 8 charges for the --SCN,
--SNO, and --N$_3$ labels, respectively. Using such more advanced
charge models is available in the CHARMM suite of programs to run
molecular dynamics
simulations.\cite{MM.dcm:2014,MM.fmdcm:2022,MM.charmm:2024} \\

\noindent
To obtain improved Lennard--Jones (LJ) parameters for probe-attached
alanine residues in lysozyme within the fMDCM framework,
methyl-substituted reference systems (CH$_3$--X) including one water
molecule were constructed. The LJ parameters for the methyl (CH$_3$)
group were taken from the CHARMM\cite{charmm:2009} force field via
CGenFF \cite{cgenff:2012}, while the probe atoms were explicitly
parametrized. For each system, a dedicated fMDCM charge model was
generated,\cite{MM.fmdcm:2022} ensuring overall charge neutrality. The
resulting electrostatic description was subsequently used to refine
the LJ parameters, yielding a balanced representation of nonbonded
interactions. For the three-atom spectroscopic probes, RKHS-based
potential energy surfaces (PESs)\cite{rabitz:1996,MM.rkhs:2017} are
available in the literature\cite{MM.n3:2022,MM.scn:2024} and have
previously been implemented in CHARMM for molecular dynamics
simulations. Since in the present work the reference cluster structure
for fitting the LJ-parameters was rigid, only the nonbonded
interactions were evaluated within CHARMM. A total of 151
configurations were randomly sampled from MD simulations for
CH$_3$--SCN, 150 configurations for CH$_3$--SNO, and 157
configurations for CH$_3$--N$_3$, respectively. Interaction energies
between water molecule and CH$_3$--X for all configurations were
computed using Gaussian\cite{g09} at the B3LYP/aug-cc-pVDZ level of
theory and were used as reference data for the LJ parameter
optimization.\\

\subsection{Molecular Dynamics}
Molecular dynamics (MD) simulations were performed using the CHARMM
package\cite{charmm:2009, MM.charmm:2024} for wild-type (WT) lysozyme
and all 51 variants in which a probe was attached to specific alanine
residues. The probes used were –SCN, –SNO, and –N$_{3}$, each bonded
to selected alanine residues at positions 41, 42, 49, 63, 73, 74, 82,
93, 97, 98, 99, 112, 129, 130, 134, 146, and 160. Each modified
residue is denoted as AlaQQX, where "QQ" is the alanine position
number and "X" represents the probe type.\\

\noindent
The initial structure of the WT lysozyme was obtained from the Protein
Data Bank (PDB ID: 1L83).\cite{1l83.pdb:1992} All simulations included
three stages: heating, equilibration, and production. The production
phase consisted of 25 ns of MD simulation at 300 K, carried out in the
\textit{NpT} ensemble. The 25-ns MD simulations of conventional MD
(cMD) simulations using CGenFF were initially performed without the
RKHS-PES and fMDCM models generated for the spectroscopic probes. In
these cMD simulations the label charges were assigned as the sum of
the atomic charges obtained from the fMDCM method, and the fitted
Lennard-Jones (LJ) parameters were also applied. The charges used for
the cMD simulations are shown in Tables \ref{sitab:table1} to
\ref{sitab:table3}.\\

\noindent
The simulations were performed in a cubic simulation box with
dimensions of $79 \times 79 \times 79$ \AA\/$^3$, using explicit TIP3P
water\cite{tip3p_1983} molecules. Bond lengths involving hydrogen
atoms were constrained using the SHAKE algorithm.\cite{shake:1977}
Nonbonded interactions, including electrostatic and van der Waals
contributions, were treated with a distance-based cutoff of 14 \AA,
with a switching function applied from 10 \AA.\cite{switch:1994} The
25 ns cMD simulations were run in the \textit{NpT} ensemble for all 17
AlaQQX modifications, after initial energy minimization, heating, and
equilibration at 300 K. As shown in Figure \ref{sifig:fig2}, the
system was re-equilibrated for 50 ps using restart files generated at
1 ns, 5 ns, 10 ns, 15 ns, 20 ns, and 25 ns from cMD
simulations. Afterwards, empirical parameters for the spectroscopic
probes were replaced by the RKHS-PES and fMDCM, followed by another 50
ps equilibration step. Finally, 0.5-ns MD simulations were conducted
for each case and probe. The fitted Lennard-Jones (LJ) parameters were
also used in these simulations. This entire procedure was repeated for
all 51 modified variants (AlaQQX).\\

\subsection{Spectroscopic Analysis}
For the protein and its subsystems, one-dimensional (1D) IR spectra
${I(\omega)}$ were obtained from the Fourier transform of the
dipole–dipole time correlation function\cite{gordon:1968,berne:2000},
\begin{equation}
  I(\omega) \propto Q(\omega) \int_0^\infty dt
  e^{i\omega t} \sum_{i=x,y,z} \left \langle \boldsymbol{\mu}_{i}(t)
  \cdot {\boldsymbol{\mu}_{i}}(0) \right \rangle
\label{eq:IR}
\end{equation}
where $\boldsymbol{\mu}(t)$ denotes the dipole moment vector of the
full protein, the modified alanine residue, or the probe moiety. A
quantum correction factor \cite{qcorr_2004},
\begin{equation}
Q(\omega) = \omega\left(1 - e^{-\beta \omega}\right),
\end{equation}
was applied to the Fourier-transformed spectra. This procedure yields
vibrational line shapes and relative intensities.\\

\noindent
To characterize the 2D-IR spectroscopy, the frequency fluctuation
correlation function (FFCF) was evaluated. For this, the frequency
trajectory $\omega(t)$ was obtained by tracking the temporal evolution
of the instantaneous normal modes
(INMs)\cite{inm:1990,MM.insulin:2020} corresponding to the C--N
stretching vibration for the --SCN probe, the N--N stretching
vibration for the --N$_3$ probe, and the N--O stretching vibration for
the --SNO probe. This INM-based analysis was performed for all 51
AlaQQX residues over $10^5$ snapshots from each
[RKHS,fMDCM]-simulation which were started from structures after 1, 5,
10, 15, 20, and 25 ns. For each snapshot, all internal degrees of
freedom of the three-atom probe were energy-minimized while the
surrounding environment was kept frozen. From $\omega(t)$ the FFCF was
computed for the fluctuation $\delta \omega(t) = \omega(t) - \langle
\omega \rangle$.  The resulting FFCF provides direct insight into
vibrational relaxation dynamics and characteristic time scales
associated with solvent--solute interactions. The FFCFs were fitted to
an empirical expression
\begin{equation}
\langle \delta \omega(t)\, \delta \omega(0) \rangle
= a_1 \cos(\gamma t)\, e^{-t/\tau_1}
+ a_2 e^{-t/\tau_2}
+ \Delta_0^2 ,
\label{eq:ffcf.fit}
\end{equation}
which enables analytical integration to obtain the corresponding line
shape function.\cite{2DIRbook-Hamm-2011} The fitting was performed
using a curve-fitting routine in the SciPy library.\cite{Virtanen2020}
In Equation \ref{eq:ffcf.fit} the parameters $a_i$, $\tau_i$,
$\gamma$, and $\Delta_0^2$ represent the amplitudes, decay time
constants, phase, and static contribution of the FFCF,
respectively. Including the first term accounts for short-time
oscillatory features in the FFCF, which are commonly attributed to
specific solvent--solute interactions and local environmental
fluctuations.\\

\section{Results}

\subsection{Performance of the fMDCM/RKHS-Models}
\begin{figure}[H]
    \centering \includegraphics[
      width=0.7\linewidth]{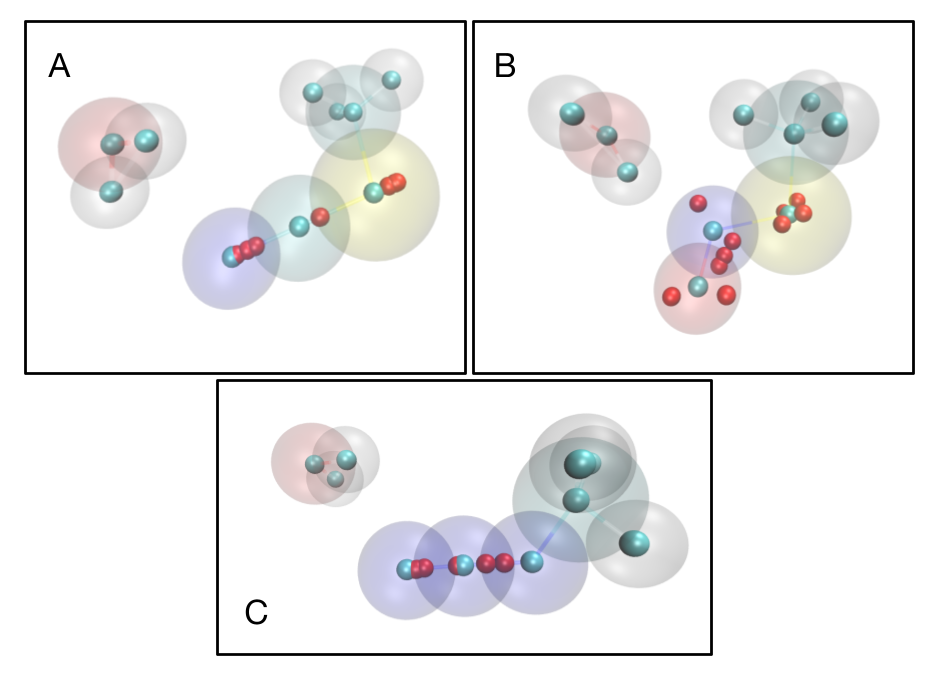}
    \caption{Three-dimensional representations of the fMDCM models,
      with atomic charges assigned according to the geometries of the
      corresponding structures with a water molecule: CH$_3$–SCN (A),
      CH$_3$–SNO (B), and CH$_3$–N$_3$ (C).}
    \label{fig:fig1}
\end{figure}

Three-dimensional representations of each probe together with the
fMDCM charges and a single water molecule are shown in Figure
\ref{fig:fig1}. The fMDCM-charges for each spectroscopic probe are
shown as red spheres, and the positions of the fMDCM-charges are the
green spheres around the center of each atom.\\

\begin{figure}[h!]
  \centering
  \includegraphics[width=0.8\textwidth]{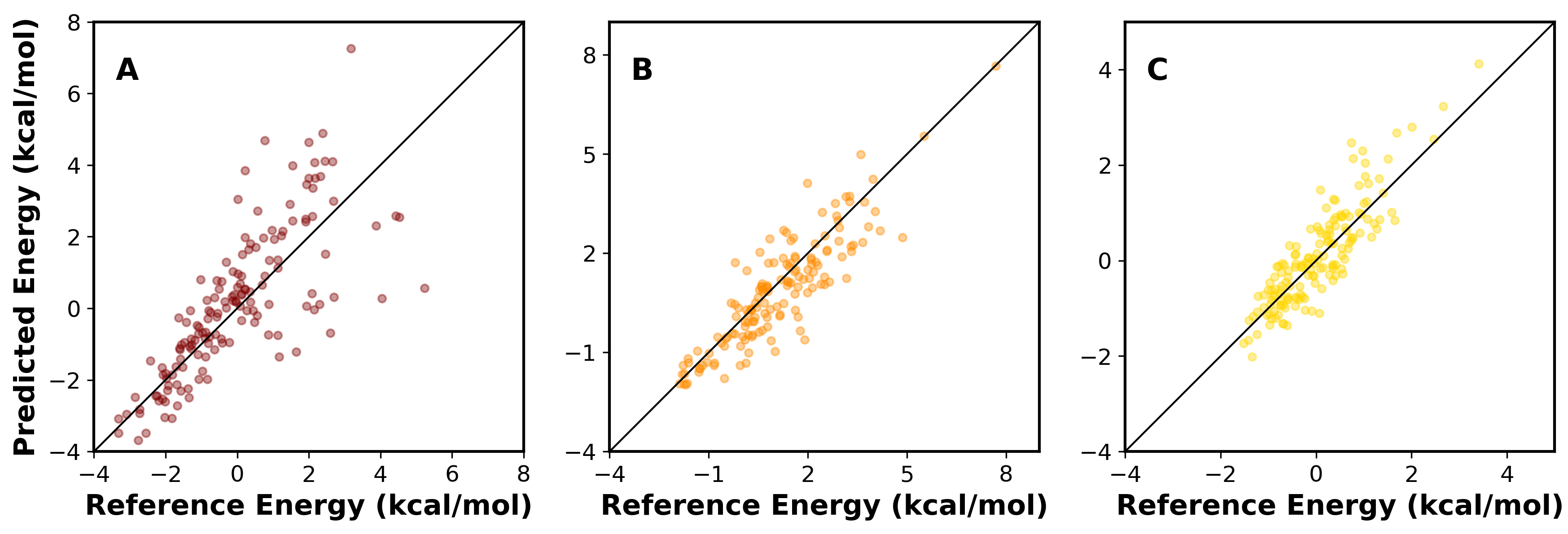}
    \caption{Correlation for Me-X model with a water molecule between
      the interaction potential ($\Delta E$) from B3LYP/aug-cc-pVDZ
      calculations with the corresponding fMDCM model interaction
      energies for each CH$_3$-X, expressed in kcal/mol. 151, 150, and
      157 snapshots are sampled to fit LJ parameters for --SCN (A),
      --SNO (B), and --N$_3$ labels (C), respectively.}
    \label{fig:fig2}
\end{figure}

\noindent
The fitting procedure was carried out using the curve fit function
from the SciPy library \cite{Virtanen2020}, employing a nonlinear
least-squares approach to minimize the residuals between CHARMM
nonbonded interaction energies and the corresponding quantum chemical
reference energies. The resulting LJ parameters for each probe are
summarized in Tables \ref{sitab:table1} to \ref{sitab:table3},
together with the atomic charges obtained from the fMDCM model. The
correlation between reference and fitted interaction energies is shown
in Figure \ref{fig:fig2}. The corresponding unfitted and fitted RMSE
values are summarized in Table \ref{sitab:table4}. \\

\noindent
In previous work, the --SCN probe has also been found to be more
challenging to represent than the --N$_3$ and --SNO
probes.\cite{cluster.MM:2026} Here, --SCN consistently exhibited
higher RMSE values than the other two probes. This may reflect a
greater sensitivity of the --SCN probe to subtle variations in its
local environment and interaction patterns, which complicates the
fitting of LJ parameters. To evaluate whether these differences
originate from the electrostatic description, the RMSE values of the
corresponding fMDCM models were examined across the ESP fitting
dataset. The RMSEs were 1.16 kcal/mol for --SCN, 1.05 kcal/mol for
--SNO, and 1.16 kcal/mol for --N$_3$. This indicates that the fMDCM
model of --SCN is of comparable accuracy compared with the other two
probes. However, when adjusting the LJ-parameters to reproduce
interaction energies with water, improving the model for --SCN is more
challenging. The resulting decrease in RMSE for all three probes, see
Table \ref{sitab:table4} demonstrates that the enhanced electrostatic
description provided by fMDCM, combined with fitted LJ-parameters,
produces a more balanced interaction model. \\

\subsection{The FFCFs}
\begin{figure}[H]
    \centering
    \includegraphics[width=1\textwidth]{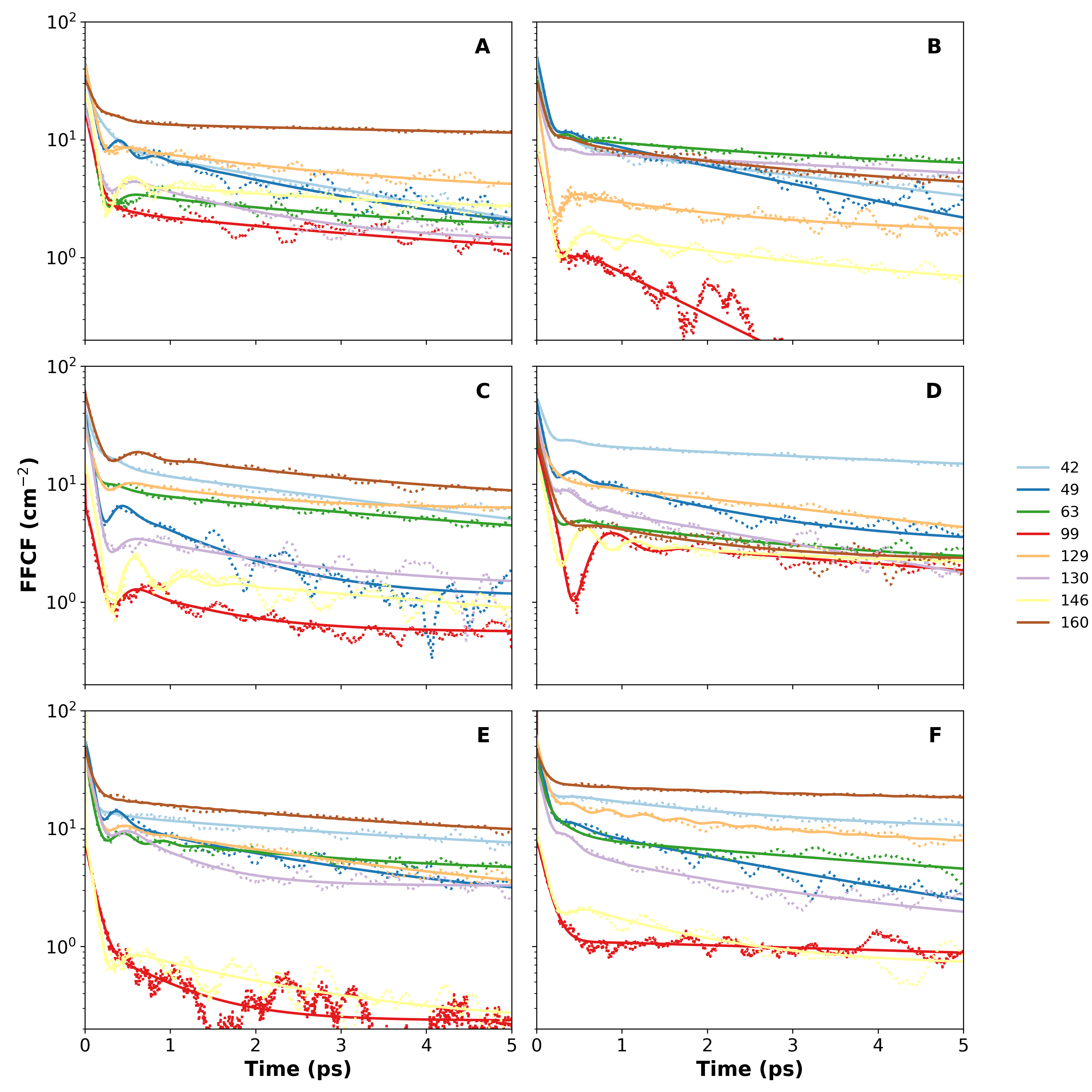}
    \caption{FFCFs for 8 Ala-positions (42, 49, 63, 99, 129, 130, 146,
      160) obtained from the INM analysis for the --SCN probe. Dashed
      lines are the raw data, and solid lines show the fits to
      Eq. \ref{eq:ffcf.fit} The $y-$axis is logarithmic. Panels A to F
      correspond to simulation windows 0.5 ns wide and the
      [RKHS,fMDCM]-simulations were started from structures after 1
      (A), 5 (B), 10 (C), 15 (D), 20 (E), and 25 (F) ns, respectively,
      of cMD simulations.}
    \label{fig:fig4}
\end{figure}

Figure \ref{fig:fig4} reports the FFCFs from the instantaneous normal
modes for eight AlaQQSCN systems, the remaining systems are shown in
Figure \ref{sifig:fig3}. Dashed lines represent the raw FFCF data,
while solid lines indicate fits to Eq. \ref{eq:ffcf.fit}. The $y-$axis
is shown on a logarithmic scale. Panels (A–F) correspond to simulation
data analyzed over 0.5 ns time windows with starting points at 1, 5,
10, 15, 20, and 25 ns, respectively.\\

\noindent
All FFCFs decay rapidly within $\sim 100$ fs but feature different
behaviours on the ps time scale. Even for a single modification,
e.g. for Ala99SCN (red trace), at different time points along the cMD
simulation, the shapes of the FFCFs change. After 1 ns (panel A) the
dynamics shows two time scales with a pronounced static component
after 5 ps whereas for the snapshot after 15 ns (panel D) a pronounced
first minimum on the sub-ps time scale appears. On the other hand, for
the snapshot after 20 ns the decay of the FFCF is bimodal on the ps
time scale, followed by oscillations and decay on the 5 ps time scale
with vanishing static contribution. This indicates that the FFCF of a
spectroscopic label is sensitive to differences in the instantaneous
conformation of the protein. Considering other locations for the
modifications corroborate this finding, see e.g. Ala129SCN or
Ala146SCN (orange, yellow), respectively. The characteristic minima in
some of the FFCFs have been observed
previously\cite{hynes:2004,li.2006,MM.cn:2013,cazade:2014} and were
linked to the interaction strength between the spectroscopic reporter
and its environment.\\

\noindent
The fits to expression \ref{eq:ffcf.fit} are of different quality. For
Ala99SCN at 15 ns (red trace) two undulations can be captured (panel
D), whereas the oscillations in panel E at 20 ns can not be described
in a realistic fashion and only the average behaviour beyond 1 ps can
be modeled. It is important to note that the irregular shapes of the
computed FFCFs are consistent with measurements. For example, similar
oscillations in the center line slope, which is related to the FFCF,
were observed for the azide anion bound in the ternary active site
complex of formate dehydrogenase and NAD$^+$.\cite{cheatum:2016} The
signal features two pronounced minima (recurrencies) and a non-zero
static contribution on the 5 ps time scale. Similarly, experiments and
simulations for the ternary azide--FDH--NAD$^+$ complex for the wild
type and two mutants at position V123 feature comparable oscillations
which differ between WT and the two mutants.\cite{pagano:2019} It
should be noted that azide interacts through electrostatic
interactions with the environment in these studies, whereas in the
present work the spectroscopic probes are covalently linked to the
Ala-residues.\\

\begin{figure}
    \centering \includegraphics[width=1\textwidth]{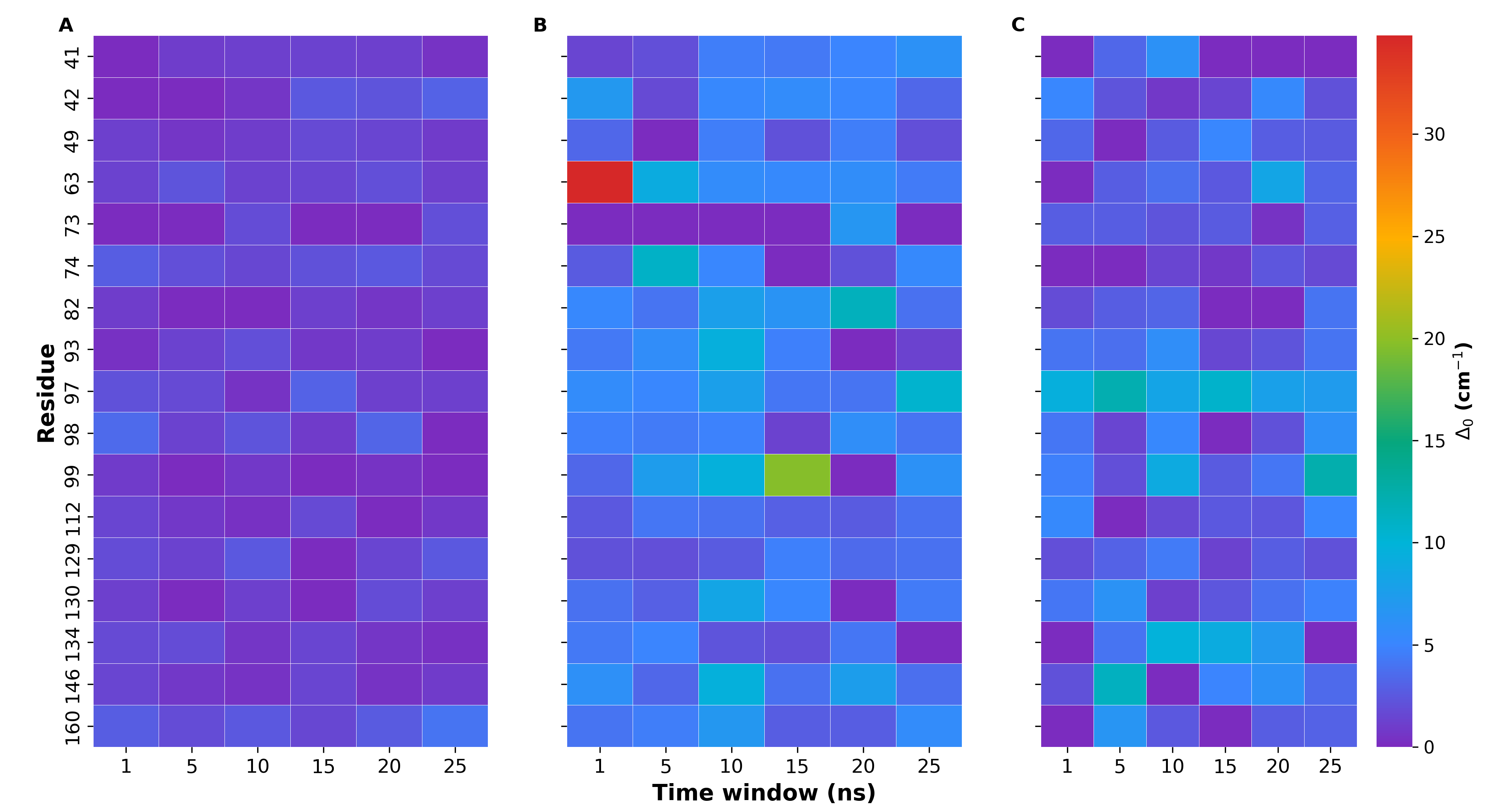}
    \caption{Heatmaps of the static
      component $\Delta_0$ (in cm$^{-1}$) obtained from fitted FFCFs
      for (A) --SCN, (B) --SNO, and (C) --N$_3$ probes across 17
      different residues for different simulation windows. Each window
      is 0.5 ns wide, and the [RKHS,fMDCM] simulations were initiated
      from structures extracted at 1, 5, 10, 15, 20, and 25 ns,
      respectively, from the cMD trajectories.}
    \label{fig:fig5}
\end{figure}

\noindent
Figure \ref{fig:fig5} presents heat maps of the static component
$\Delta_0$ obtained from the fits of the FFCFs for --SCN, --SNO, and
--N$_3$ probes. For each residue, the heat maps include data collected
from the six simulation windows, allowing a direct comparison of the
static contributions across the trajectories. For the --SCN probe
(Figure \ref{fig:fig5}A), the largest static components are observed
for Ala63SCN, Ala74SCN, and Ala160SCN. At these positions, the static
component remains above 1 cm$^{-2}$ across all simulation windows. In
particular, Ala160SCN exhibits consistently high static contributions,
with $\Delta_0-$values ranging from 1.5 to 4.0 cm$^{-1}$ ($\Delta_0^2
= 2.24$ to 16.13 cm$^{-2}$) across the simulation windows. These
results indicate that the frequency fluctuations at these sites
contain a substantial static contribution that persists across the
different stages of the trajectories. The --SNO probe exhibits the
largest static components among the three probes, with substantially
higher $\Delta_0^2$ values observed at many residue positions and
across multiple simulation windows (Figure \ref{fig:fig5}B). In
contrast to --SCN, where only a few residues exhibit consistently high
static contributions, elevated $\Delta_0^2$ values for --SNO are
observed at a larger number of positions. The --N$_3$ probe also
exhibits relatively large static contributions compared with --SCN
(Figure \ref{fig:fig5}C). In particular, Ala97N$_3$, Ala99N$_3$, and
Ala146N$_3$ display $\Delta_0^2$ values exceeding 100 cm$^{-2}$ in
some simulation windows. Thus, the overall magnitude of the static
contribution follows the general trend of --SNO $>$ --N$_3$ $>$ --SCN,
although substantial residue- and window-dependent variations are
observed for all three probe.\\

\noindent
Decay times, $\tau_1$ and $\tau_2$ (in ps, upper and lower rows,
respectively), obtained from fitting the FFCF to
\ref{eq:ffcf.fit} for the INM frequencies of all AlaQQX residues in
lysozyme are listed in Tables \ref{sitab:table5}, \ref{sitab:table6},
and \ref{sitab:table7}. The decay times show a different degree of
variation across the simulation windows. As an example, for
Ala146N$_3$, $\tau_2 = 0.46$ ps in the 5 ns simulation window, for
which $\Delta_0 = 11.35$ cm$^{-1}$ ($\Delta_0^2 = 128.7$ cm$^{-2}$) is
particularly large. In contrast, simulation windows with lower static
components can exhibit larger $\tau_2$ values. For instance, in the 10
ns simulation windows of Ala146N$_3$, $\tau_2$ increases to 4.52 ps,
indicating that the timescale of the slower frequency fluctuations can
depend substantially on the simulation window, corresponding to local,
rapidly decorrelating motions.\\

\noindent
It is also of interest to compare these characteristics with
measurements. Finite values of $\Delta_0$ have been associated with
residual dynamics which has not decayed on the time scale of the
analysis/simulation. Experimentally measured static components for
--SCN attached to calmodulin and PYP range from 2 to 4 cm$^{-1}$ and 3
to 5 cm$^{-1}$,
respectively.\cite{bredenbeck.2:2020,bredenbeck.3:2020}. For free and
ligated CN-labelled P450cam $\Delta_0-$values of 1 to 4 cm$^{-1}$),
were reported,\cite{thielges:2021} whereas nitrile probes
(cyanophenylalanine) in HP35 and in S-peptide feature values of 2.2
cm$^{-1}$ and 2.4 cm$^{-1}$,
respectively.\cite{Chung.hp352dir.pnas.2011,bagchi:2012} Static
inhomogeneous components were also observed in the spectral diffusion
of azide probes in a peptide bound to a protein.\cite{hamm.jpcb.2012}
Finally, previous simulations of cyano-substituted phenol in lysozyme
reported static components of $\sim 2$ cm$^{-1}$, comparable with the
present findings.\cite{MM.lys:2017} In bulk isotropic solvents, local
hydrogen-bonding and solvation dynamics fully average out on the
picosecond timescale, driving the static component to zero ($\Delta_0
\sim 0$ cm$^{-1}$). This was, for example, observed for CN$^-$ in
water\cite{Kozinski07p5,MM.cn:2013} and for bulk thiocyanate
(SCN$^-$).\cite{Renzhe:2015} Similarly, time scales $\tau_1 \sim 0.1$
ps and $\tau_2 \sim 2$ ps from the computations are broadly consistent
with measurements, although depending on the system considered,
variations in $\tau_2$ can be
substantial.\cite{roberts:2009,yuan:2015}\\

\subsection{IR Spectroscopy of the Probes}

\begin{figure}[h!]
    \centering
    \includegraphics[width=1\textwidth]{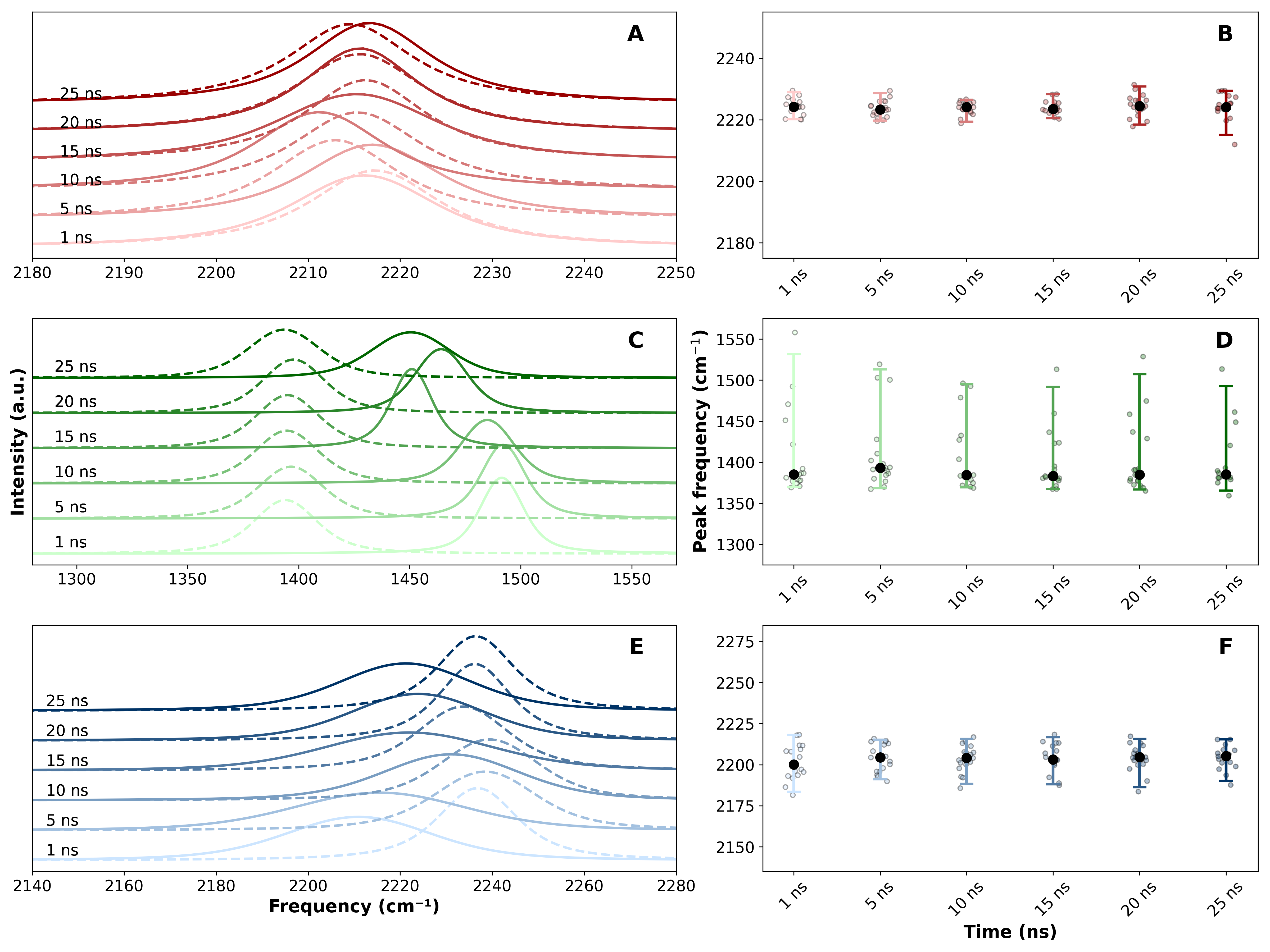}
    \caption{Left panel: 1D-IR line shape from the FFCF for Ala41X
      (dashed line) and Ala97X (solid line) in lysozyme for all labels
      in time with X = --SCN (A), --SNO (C), and --N$_3$ (E) from top
      to bottom. The ranges of conformationally induced frequency
      shifts are $\omega_{\rm SCN} \in [2212,2217]$, $\omega_{\rm SNO}
      \in [1393,1492]$, and $\omega_{\rm N_3} \in [2210,2240]$
      cm$^{-1}$. For each trace, the FFCF for 0.5 ns of MD data
      preceding the time stamp indicated was computed and the Fourier
      Transform was determined to obtain the 1D-IR lineshape. Right
      panel: Distribution of peak frequencies from 1D-IR spectra
      calculated dipole moments for C--N stretching vibration of the
      --SCN probe (B), the N--O stretching vibration of the --SNO
      probe (D), and the N--N stretching vibration of the --N$_3$
      probe (F), for all 17 labelled Ala-residues in lysozyme from all
      [RKHS,fMDCM]-simulation windows of 1, 5, 10, 15, 20, and 25 ns.}
    \label{fig:fig3}
\end{figure}

\noindent
To investigate the local dynamics and vibrational spectroscopy of the
C–N, N–O, and N–N stretching modes of the –SCN, –SNO, and –N$_3$
probes, respectively, one-dimensional IR (1D-IR) line shapes were
obtained from the FFCFs as well as from subsystem dipole moments for
all modified AlaQQX residues. In the left panels of Figure
\ref{fig:fig3}, Ala41X and Ala97X are selected as representative
examples due to their distinct peak positions. Figures
\ref{fig:fig3}A/C/E illustrate the temporal evolution of band
position, shape, and width, reflecting changes in the local probe
environment. As shown in Figure \ref{fig:fig3}A, the spectral
differences between Ala41SCN and Ala97SCN are comparatively small for
the --SCN probe, with frequency ranges of 2213–2217 cm$^{-1}$ for
Ala41SCN and 2211–2217 cm$^{-1}$ for Ala97SCN across different time
windows. In contrast, pronounced variations in peak position are
observed for the --SNO and --N$_3$ probes, indicating a stronger
sensitivity of these vibrational modes to changes in the local
environment. For the --SNO probe, the site-dependent differences are
more pronounced. Ala97SNO exhibits peak positions ranging from 1450 to
1492 cm$^{-1}$ across the different simulation windows, whereas
Ala41SNO shows a much narrower range of 1393–1398 cm$^{-1}$. In
addition to the larger variation observed for Ala97SNO, the
characteristic frequencies of the two sites differ by more than 50
cm$^{-1}$. This substantial site-dependent frequency difference may
arise from differences in the local environments of the two
probes. Ala41SNO remains in a hydrated environment across the
simulation windows, whereas Ala97SNO is located in a predominantly dry
region. The presence or absence of surrounding water molecules can
modify the local electrostatic environment and hydrogen-bonding
interactions of the probe, thereby shifting its vibrational
frequency. The broader frequency range observed for Ala97SNO may
additionally reflect greater sensitivity to local structural
fluctuations in the less hydrated environment. These observations
suggest that both the hydration state and the local structural
environment contribute to the site-dependent spectral response of the
--SNO probe.\\

\noindent
In Figure \ref{fig:fig3}, the right panels show frequency fluctuation
bars of peak frequency distributions extracted from the IR spectra for
the --SCN (B), --SNO (D), and --N$_3$ (F) probes, obtained from
subsystem dipole moments. These spectra were calculated using the
atomic coordinates and charges of the probe, corresponding to the C–N,
N–O, and N–N stretching vibrations, respectively, across 17 different
probe-attached residues. For each spectrum, the peak frequency was
identified as the frequency corresponding to the maximum intensity
within the selected spectral region (colored dots). These peak
frequencies were then collected for all spectra within each time
window, providing a distribution that characterizes the variability of
the observed vibrational frequency. For each time window, the
individual peak frequencies are shown together with their median
values (black dot) and central 95\% intervals, defined by the 2.5th
and 97.5th percentiles of the peak-frequency distributions.\\

\noindent
A related position-dependent spectroscopic behavior was observed in a
previous work on insulin.\cite{MM.insulin:2020} The same --CO probe
exhibits substantially different absorption frequencies depending on
its location within the protein, despite having the same underlying
bonded potential. In particular, probes located at the dimerization
interface exhibit distinct spectral shifts upon dimer
formation.\cite{MM.insulin:2020} This observation demonstrates that
the frequency of an otherwise identical vibrational probe can be
strongly modulated by its local protein environment, highlighting the
sensitivity of infrared probes to the underlying structural and
dynamical heterogeneity.\\

\subsection{Dynamics of --SCN-Functionalized Ala-Residues}
Next, the coupling between the high-frequency mode of the
spectroscopic probe and the low-frequency motions of the protein was
analyzed. For this, the protein-IR spectrum was determined from the
total dipole moment correlation function and compared with the
lineshape obtained from Fourier transformation of the label-specific
FFCF. FFCFs for all modification sites and for all three labels were
determined. However, only particularly notable cases are discussed in
detail in the following.\\

\noindent
For --SCN-labelled Alanine residues, the analysis focused on positions
99 and 146. As shown in Figure \ref{fig:fig4}, the FFCFs for AlaQQSCN
residues at positions 99 and 146 display pronounced variations across
different time windows. In these cases, oscillatory features in the
FFCF are clearly distinguishable in certain simulation windows, while
the oscillatory character diminishes or disappears in others. To
elucidate the origin of this behavior, Fourier transforms of the FFCFs
were performed to examine the corresponding frequency spectrum and
potential coupling between the spectroscopic probe and the protein
modes.\\

\begin{figure}[h!]
    \centering \includegraphics[width=0.8\textwidth]{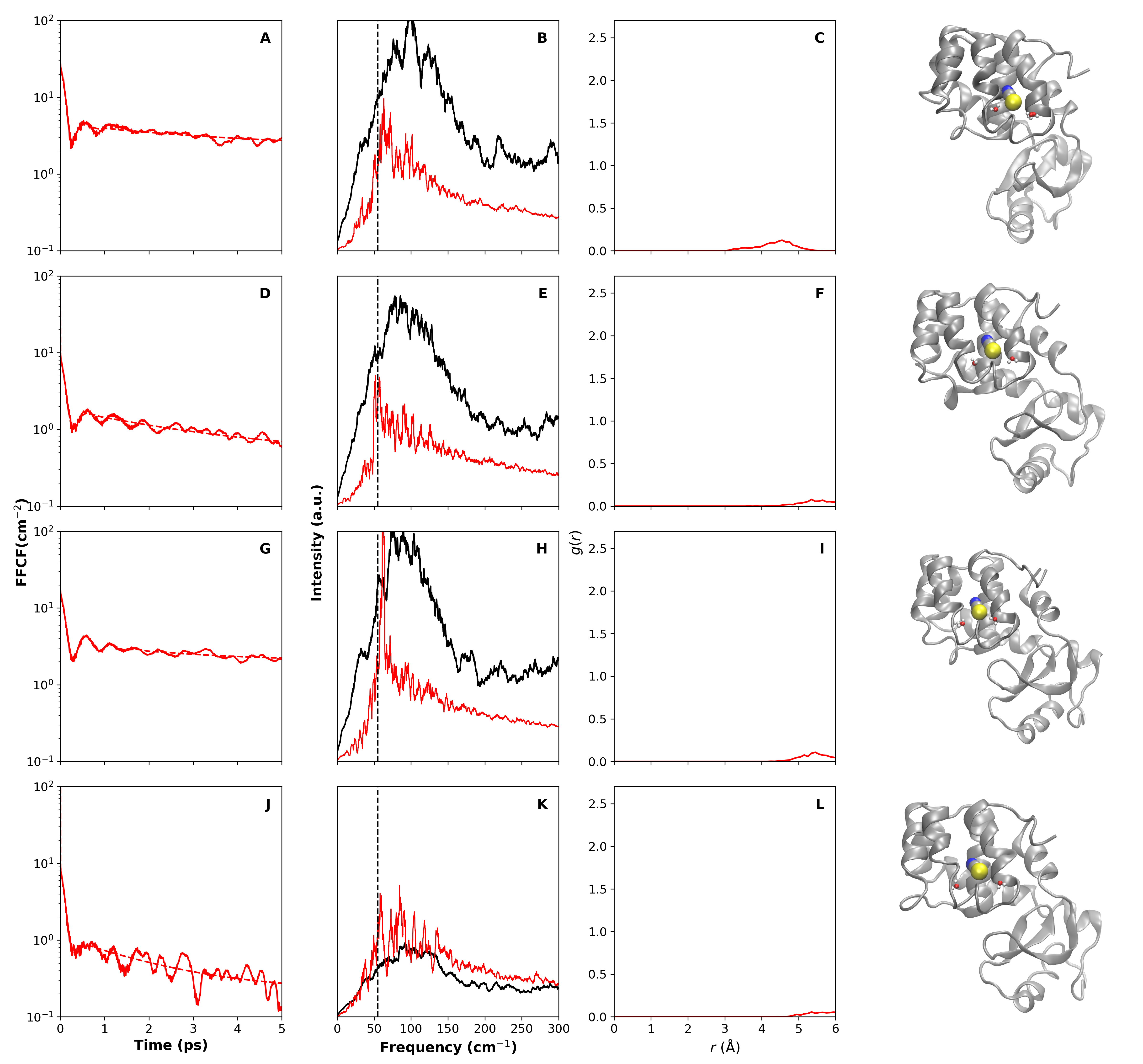}
    \caption{Left-hand panels: FFCFs for Ala146SCN obtained from the
      INM analysis for the --SCN probe. Middle panel: For the --SCN
      probe, protein with probe IR spectra from dipole moment (black
      line), and the FFT of the FFCF (red line) is also shown at
      low-frequency for Ala146SCN. Black dashed vertical lines mark
      the approximate FFT peak positions and serve as a visual guide
      for comparison with the corresponding features in the calculated
      protein with probe IR spectra. Right-hand panels: Radial
      distribution functions (RDFs) between the oxygen atoms of water
      molecules and the central atom of the –SCN probe. In addition, a
      three-dimensional structural representation of the probe
      location within the protein, including the surrounding residues
      and nearby water molecules, is shown in the far-right
      panels. From top to bottom, each row corresponds to
      [RKHS,fMDCM]-simulation windows of 1, 5, 15, and 20,
      respectively. The analyzed data window was 0.5 ns wide.}
      \label{fig:fig6}
\end{figure}

\noindent
The left-hand panels of Figure \ref{fig:fig6} present the FFCFs for
Ala146SCN obtained from instantaneous normal mode (INM) analysis of
the --SCN probe. From top to bottom, each row corresponds to
simulations initialized after 1, 5, 15, and 20 ns of cMD simulations,
respectively. The FFT of the FFCFs (red), together with the IR
spectrum of the full protein (black), is shown in the middle
panels. The right-hand panels display the radial distribution
functions (RDFs) between the oxygen atoms of water molecules and the
central atom of the --SCN probe. Starting from the 1 ns window, the
FFCF develops a more pronounced oscillatory behavior, which becomes
particularly evident in the 15 ns window, indicating relatively
restricted environmental fluctuations around the probe, consistent
with a more crowded and less hydrated local environment. In these
windows, the decay of the FFCF is slower compared to the 20 ns window,
and clear oscillatory modulations are observed, see Figures
\ref{fig:fig6}A/D/G. Analysis of the corresponding Fourier transforms
in Figures \ref{fig:fig6}B/E/H reveals a dominant oscillation
frequency at $\approx 55$ cm$^{-1}$. The pronounced oscillatory
behavior observed in Figures \ref{fig:fig6}A/D/G suggests that the
probes may also contribute to the overall protein IR response. No
pronounced feature is apparent around 55 cm$^{-1}$ in Figures
\ref{fig:fig6}B/E, though. However, the protein IR spectrum in Figure
\ref{fig:fig6}H (black line) exhibits a weak peak in this spectral
region, suggesting that the oscillations of the probes may contribute,
albeit weakly, to the protein IR response. In contrast, the 20 ns
window shows a faster decay of the FFCF, and the corresponding FFT
spectrum indicates a reduction in relative intensity in the
low-frequency region, suggesting a loss of coherent low-frequency
environmental modes.\\

\begin{figure}[h!]
    \centering
    \includegraphics[width=0.8\textwidth]{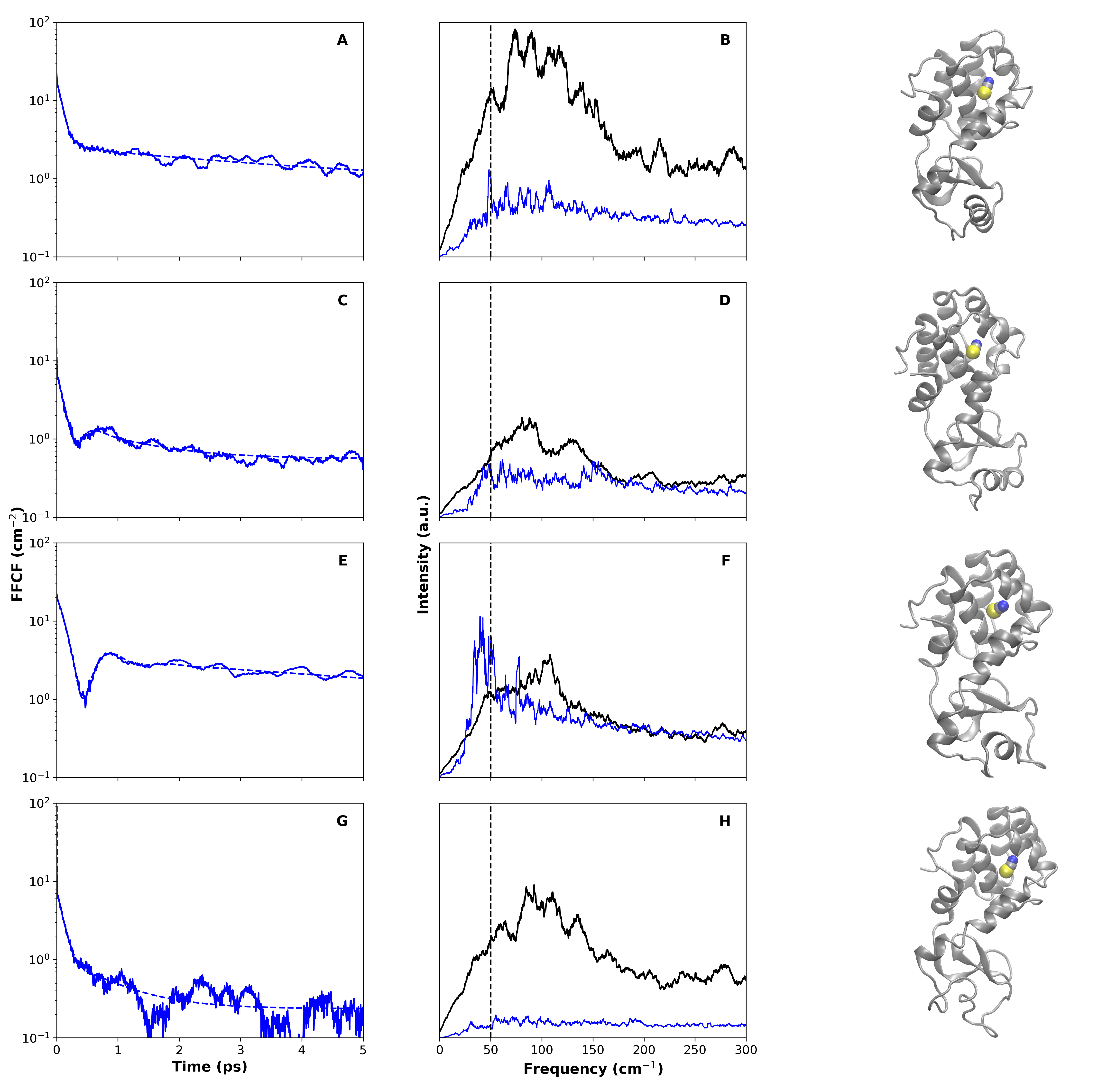}
    \caption{Left-hand panels: FFCFs for Ala99SCN obtained from the
      INM analysis for the --SCN probe. Right-hand panel: For the –SCN
      probe, protein with probe IR spectra from dipole moment (black
      line), and the FFT of the FFCF (blue line) is also shown at
      low-frequency for Ala99SCN. Black dashed vertical lines mark the
      approximate FFT peak positions and serve as a visual guide for
      comparison with the corresponding features in the calculated
      protein with probe IR spectra. In addition, a three-dimensional
      structural representation of the probe location within the
      protein, including the surrounding residues and nearby water
      molecules, is shown in the far-right panels. From top to bottom,
      each row corresponds to [RKHS,fMDCM]-simulation windows of 1,
      10, 15, and 20, respectively. The analyzed data window was 0.5
      ns wide.}
      \label{fig:fig7}
\end{figure}

\noindent
A similar analysis was performed for Ala99SCN. In this case, the local
environment of the probe remains essentially dehydrated throughout the
windows, with no water molecules observed in its immediate vicinity up
to 6 \AA. Despite the absence of solvent interactions, a pronounced
frequency, approximately 50 cm$^{-1}$, in the FFCF is observed only in
the 15 ns window (Figure \ref{fig:fig7}F), whereas in other time
windows the oscillatory character diminishes. Interestingly, at the 1
ns and 10 ns windows, distinct peaks emerge starting from
approximately 50 cm$^{-1}$, accompanied by a broader distribution of
frequencies. At the 15 ns window, both the peaks and oscillatory
features of the FFCF become more pronounced, appearing at frequencies
below 50 cm$^{-1}$. In contrast, at the 20 ns window, a significantly
faster decay is observed compared to the other simulations, and the
FFT of the FFCF shows a clear loss of peak intensity in the
low-frequency pattern.\\

\noindent
To further investigate the role of protein dynamics, principal
component analysis (PCA) was performed separately for each of six
consecutive time windows using the distances of heavy protein atoms
within 7 \AA\/ of the center of mass of the –SCN probe (excluding
Ala99) for Ala99SCN, see Figure \ref{sifig:fig4}. The PCA projections
show a broadly cloud-like distribution in all windows except the 25 ns
window, indicating a flexible and heterogeneous local protein
environment. In Figure \ref{sifig:fig4}D, for the 15 ns window, the
PCA distribution becomes more compact and shifts toward relatively
denser sampling region, accompanied by an increased radius of gyration
of approximately 6.5 \AA, suggesting a transient expansion of the
local protein environment. Additionally, this window exhibits more
pronounced oscillatory behavior in the FFCF together with enhanced
spectral features, indicating a stronger coupling between local
protein rearrangements and vibrational frequency fluctuations.\\

\begin{figure}[h!]
    \centering
    \includegraphics[width=1\linewidth]{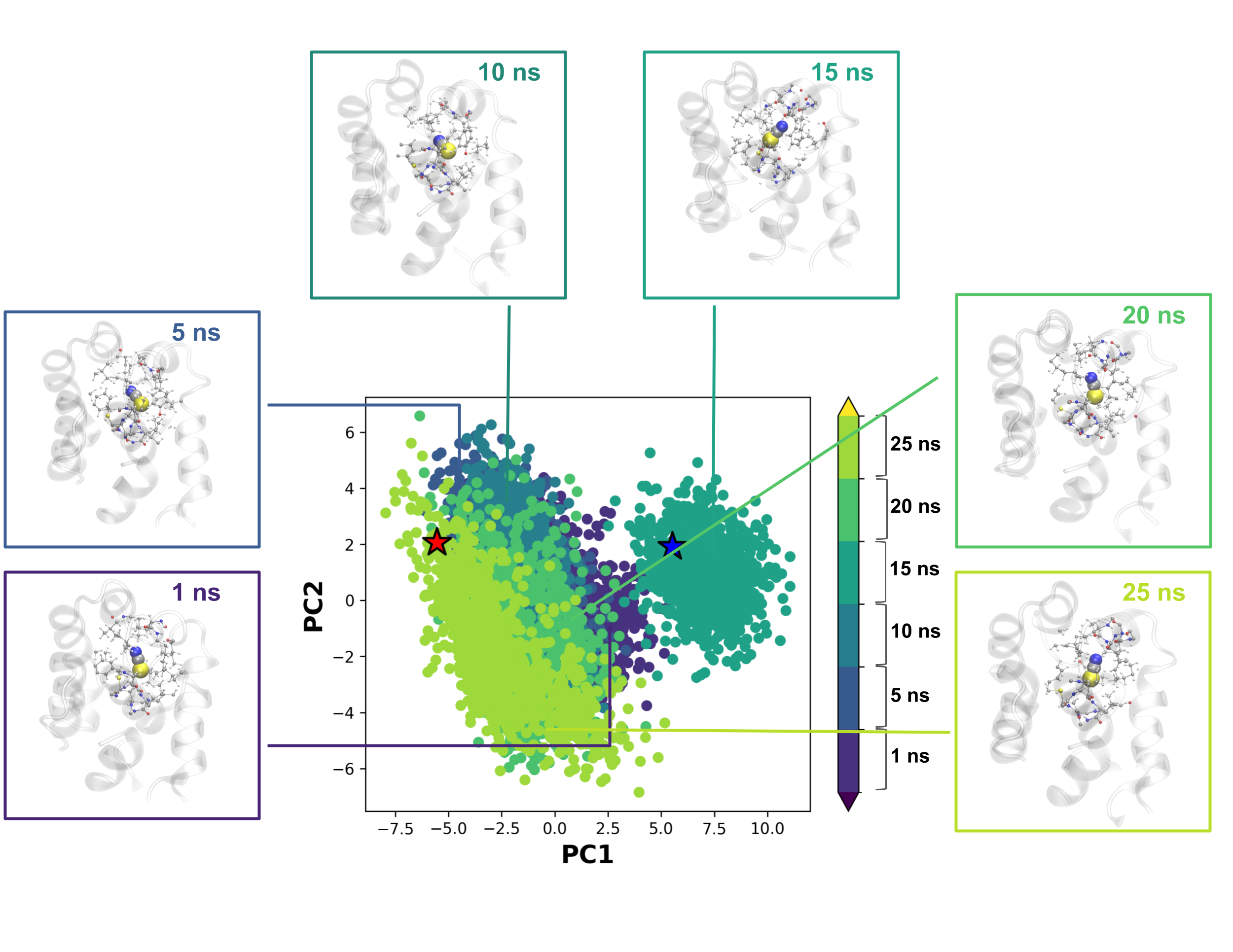}
    \caption{PCA of protein--SCN distance features for Ala99SCN
      computed using distances between the --SCN probe and individual
      heavy atoms of protein residues within 7 \AA\/ of the
      probe. Distances were calculated for each atom in the selected
      region and projected onto the first two principal
      components. Points are colored by simulation time, illustrating
      the temporal evolution of conformational sampling and
      transitions between distinct probe–protein interaction
      states. Both stars are represented the frames having similar PC2
      values but opposite PC1 values. Frames were saved every 0.5 ps
      and aligned with respect to the protein backbone.}
    \label{sifig:fig5}
\end{figure}

\noindent
In contrast, for the 20 and 25 ns windows, the PCA distributions
separate into more distinct clusters, reflecting slower and more
heterogeneous conformational transitions, see Figures
\ref{sifig:fig4}E/F. This structural change is accompanied by a
modification of the FFCF dynamics in Figure \ref{fig:fig7} at 20 ns,
where a faster decay is observed, consistent with a rearrangement of
the probe environment and altered local fluctuations. To complement
this analysis, PCA of protein–SCN distance features for Ala99SCN was
performed using distances between the SCN probe and heavy atoms of
protein within 7 \AA, see Figure \ref{sifig:fig5}. The resulting
trajectories from all time windows were projected onto the first two
principal components using frames saved every 0.5 ps and aligned to
the protein backbone. Coloring by simulation time reveals the temporal
evolution of conformational sampling and transitions between distinct
probe–protein interaction states. In the 15 ns window, a pronounced
deviation along PC1 in PCA space is observed relative to the other
windows, indicating a transient local rearrangement of the probe
environment. Such behavior is not observed in the remaining windows,
which display more continuous sampling patterns consistent with the
trends observed in Figure \ref{sifig:fig4}.\\

\noindent
To further characterize this transition, representative conformations
were selected from regions of the PCA space with comparable PC2 values
but opposite extremes along PC1, see Figure \ref{sifig:fig5}. Two
frames corresponding to these regions were extracted, RMSD-aligned
using the protein backbone, and visualized in three dimensions. The
probe is shown in van der Waals representation, the protein in
NewCartoon representation, and heavy atoms of residues surrounding the
probe in CPK representation. These structures illustrate distinct
probe–protein interaction states located on opposite sides of PC1
while maintaining comparable PC2 values. Protein backbone RMSD between
the states and protein within 7 \AA\/ of the center of mass of --SCN
probe RMSD between the states are 2.02 \AA\/ and 1.53 \AA\/,
respectively, see Figure \ref{sifig:fig6}. \\

\subsection{Dynamics of --N$_3$- and --SNO-Functionalized Ala-Residues}
While the --SCN probe exhibits the most pronounced oscillatory and
strongly time-dependent patterns, similar features can also be
observed for the --SNO and --N$_3$ probes across different residues,
although with lower amplitudes. In particular, several N$_3$-labeled
residues, including Ala98N$_3$, Ala129N$_3$, and Ala160N$_3$, display
oscillatory patterns, see Figures \ref{sifig:fig3-N3-1} and
\ref{sifig:fig3-N3-2}. This suggests the presence of similar dynamic
behavior, albeit less pronounced than that observed for the
corresponding SCN-labeled residues. To further investigate these
behaviors, representative cases exhibiting either oscillatory features
or distinct decay characteristics were selected, namely Ala146SNO and
Ala82N$_3$. A similar analysis as for Ala99SCN and Ala146SCN was
carried out.\\

\begin{figure}[h!]
    \centering \includegraphics[width=0.8\textwidth]{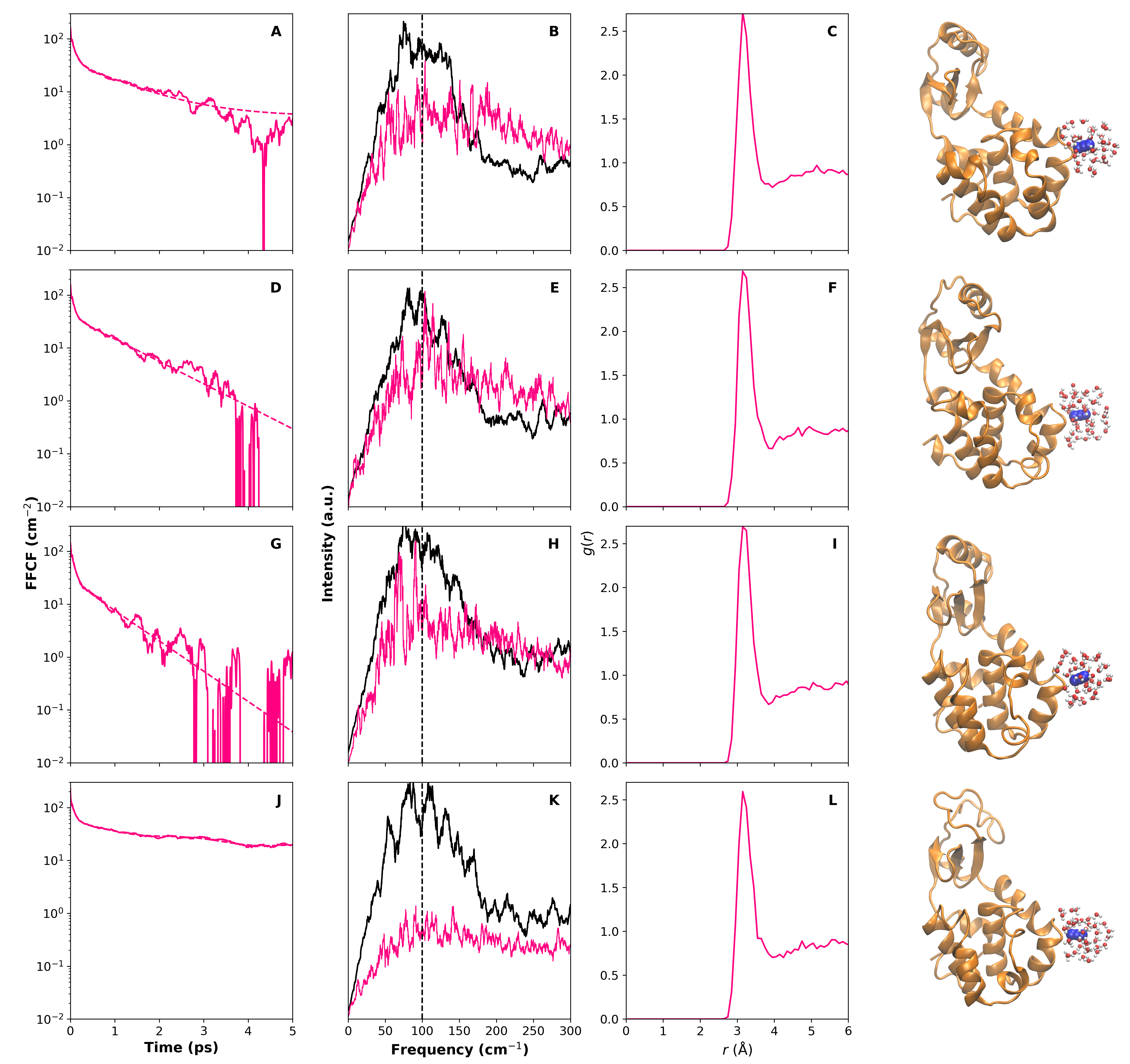}
    \caption{Left-hand panels: FFCFs for Ala82N$_3$ obtained from the
      INM analysis for the --N$_3$ probe. Middle panel: For the –N$_3$
      probe, protein with probe IR spectra from dipole moment (black
      line), and the FFT of the FFCF (colored line) is also shown at
      low-frequency for Ala82N$_3$. Black dashed vertical lines mark
      the approximate FFT peak positions and serve as a visual guide
      for comparison with the corresponding features in the calculated
      protein with probe IR spectra. Right-hand panels: Radial
      distribution functions (RDFs) between the oxygen atoms of water
      molecules and the central atom of the --N$_3$ probe. From top to
      bottom, each row corresponds to [RKHS,fMDCM]-simulation windows
      of 1, 15, 20, and 25, respectively. The analyzed data window was
      0.5 ns wide. In addition, a three-dimensional structural
      representation of the probe location within the protein,
      including the surrounding residues and nearby water molecules,
      is shown in the far-right panels.}
      \label{fig:fig9}
\end{figure}

\noindent
For the --N$_3$ probe, the selected residue represents a distinct
dynamic behavior characterized by a decay profile rather than the
pronounced oscillatory features observed for the --SCN probe. For
Ala82N$_3$, a similar analysis was performed for the 1, 15, 20, and 25
ns simulation windows, each evaluated over a 0.5 ns data segment in
Figure \ref{fig:fig9}. In contrast to the oscillatory behavior
observed for some --SCN and --SNO cases, the N--N stretching mode
primarily exhibits a rapidly decaying FFCF, indicating fast loss of
frequency correlation for Ala82N$_3$. As evident from the right-hand
panels, including both the structural representations and the radial
distribution functions, the probe remains well hydrated across all
time windows. Notably, in Figures \ref{fig:fig9}B/E, a more pronounced
spectral feature is observed in the 50--100 cm$^{-1}$ range compared
to other initial-state windows. This behavior might be associated with
short-time oscillatory components in the FFCF within the 2--5 ps
regime, as illustrated in Figures \ref{fig:fig9}D/G.\\

\begin{figure}[h!]
    \centering \includegraphics[width=0.8\textwidth]{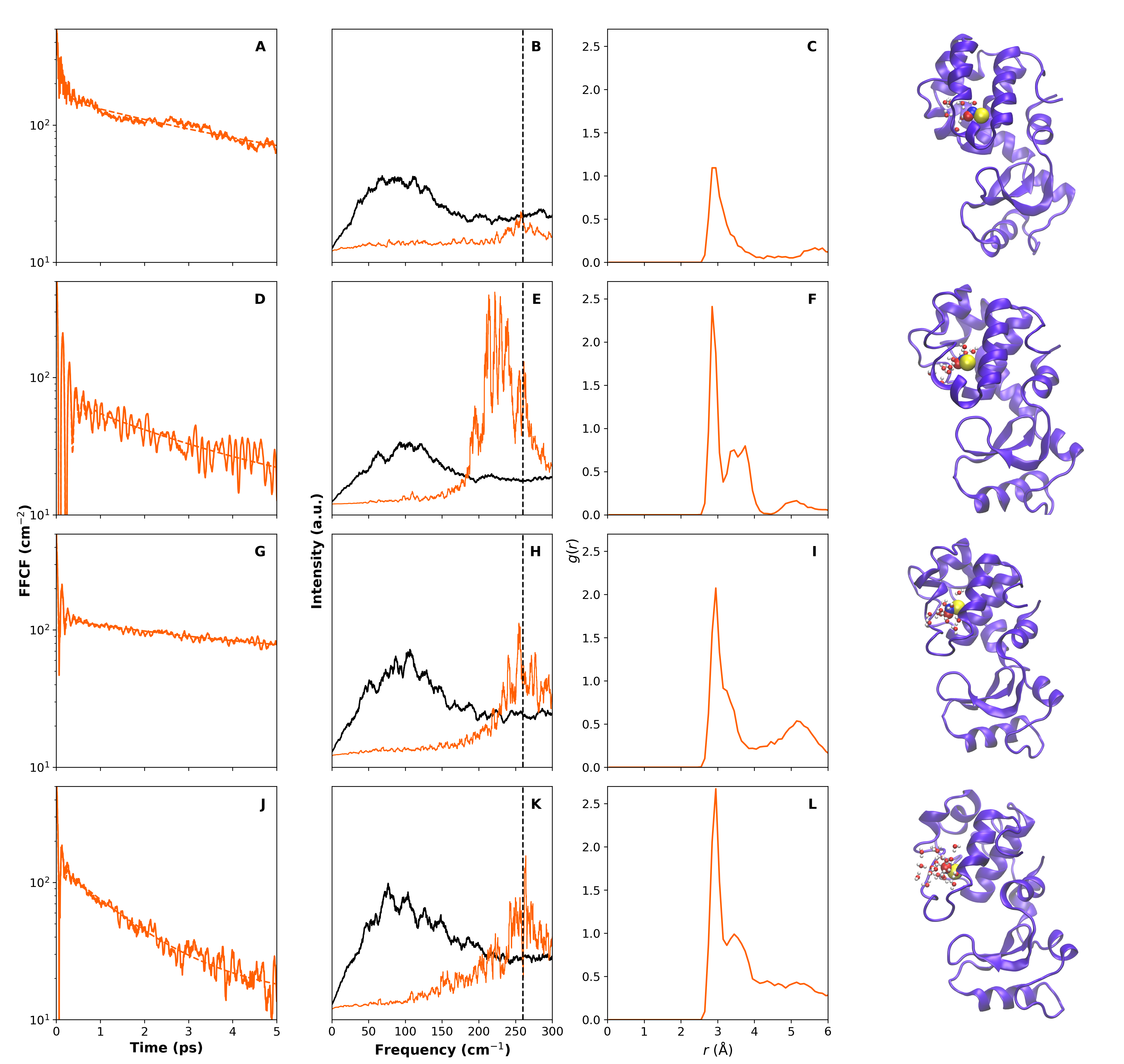}
    \caption{Left-hand panels: FFCFs for Ala146SNO obtained from the
      INM analysis for the --SNO probe. Middle panel: For the –SNO
      probe, protein with probe IR spectra from dipole moment (black
      line), and the FFT of the FFCF (colored line) is also shown at
      low-frequency for Ala146SNO. Black dashed vertical lines mark
      the approximate FFT peak positions and serve as a visual guide
      for comparison with the corresponding features in the calculated
      protein with probe IR spectra. Right-hand panels: Radial
      distribution functions (RDFs) between the oxygen atoms of water
      molecules and the central atom of the --SNO probe. In addition,
      a three-dimensional structural representation of the probe
      location within the protein, including the surrounding residues
      and nearby water molecules, is shown in the far-right
      panels. From top to bottom, each row corresponds to
      [RKHS,fMDCM]-simulation windows of 1, 5, 20, and 25,
      respectively. The analyzed data window was 0.5 ns wide.}
      \label{fig:fig8}
\end{figure}

\noindent
Figure \ref{fig:fig8} reports the results for Ala146SNO. From top to
bottom, the rows correspond to simulation windows of 1, 10, 15, and 20
ns, and the windows of 1, 5, 20, and 25 ns, respectively, each
analyzed over 0.5 ns segments from simulations including both
corresponding fMDCM and PES descriptions. As shown in the right-hand
panels of Figure \ref{fig:fig8}C/F/I/L, water molecules are present in
the local environment of the probe in all time windows. For Ala146SNO,
the FFCF exhibits pronounced oscillating behavior, particularly in the
5 ns window in Figure \ref{fig:fig8}. This feature is also reflected
in the corresponding FFT of the FFCF, where a distinct frequency
component in the range of 200--260 cm$^{-1}$ is observed, see Figure
\ref{fig:fig8}E. Similar frequency contributions persist in the 20 and
25 ns windows (Figure \ref{fig:fig8}H/K, respectively) at 260
cm$^{-1}$, which can be attributed to oscillatory behavior of the FFCF
at short times (0--0.5 ps) in Figures \ref{fig:fig8}G/J.\\

\section{Discussion and Conclusion}
In this study, three vibrational probes, --SCN, --SNO, and --N$_3$,
were used to characterize site-specific conformational dynamics in
lysozyme. Each probe was introduced at 17 alanine sites spanning
distinct environments, enabling a direct comparison of their
sensitivity to variations in hydration, local structural packing, and
conformational flexibility. Spectroscopic properties were evaluated
for both hydrated and relatively dry environments, with particular
emphasis on low-frequency vibrational modes and their relationship to
the surrounding molecular dynamics, spectroscopy, and structure. The
resulting frequency fluctuations were characterized through
Fourier-transform analysis of the FFCFs to assess their contributions
to the overall protein spectral response, while calculated 1D-IR
spectra based on the probe dipole moments were used to further
evaluate the spectroscopic signatures associated with each probe and
labeling site. These complementary analyses revealed probe- and
site-dependent differences in local dynamics and provided insight into
how variations in structural and hydration environments are reflected
in the vibrational response of the protein. \\

\noindent
Experimental studies using --SCN-labelled proteins have demonstrated
that the vibrational response is coupled to local solvation and
structural fluctuations. Previous 2D-IR measurements of SCN-labelled
calmodulin demonstrated that the probe can resolve site-specific
dynamics, solvent exposure, and conformational changes through
FFCF.\cite{bredenbeck.2:2020} Especially, differences in --SCN
dynamics were observed between hydrated and buried environments,
showing that the vibrational response is sensitive to local
hydration. In addition, measurements of SCN-labelled photoactive
yellow protein in H$_2$O and D$_2$O revealed that the vibrational
lifetime of the --SCN oscillator strongly depends on solvent
accessibility, with water-exposed labels exhibiting enhanced
vibrational relaxation.\cite{bredenbeck.3:2020} These observations
indicate that changes in hydration surrounding the probe directly
influence its spectroscopic behavior. In the light of these
experimental results, the current results from the simulations show
that --SCN recurrence events are associated with reduced local water
occupancy.\\

\noindent
The decaying FFCFs observed in Figure \ref{fig:fig9}, together with
the various oscillatory FFCF behaviors shown in Figures
\ref{sifig:fig3-N3-1} and \ref{sifig:fig3-N3-2}, highlight the diverse
vibrational dynamics captured by --N$_3$. Previous studies showed that
azide vibrational frequencies are highly sensitive to local
hydrogen-bonding interactions and electrostatic
fluctuations.\cite{cho:2015,cho.2008-3} Since the azide directly
responds to changes in water dynamics, hydrogen-bond geometry, and
local electric fields, its frequency fluctuations contain substantial
contributions from fast solvent motions in addition to slower protein
structural fluctuations. For Ala82N$_3$ see Figure \ref{fig:fig9}, the
probe remains hydrated throughout the trajectories; however, its FFCF
exhibits a fast decay pattern, suggesting that rapid local hydration
and electrostatic fluctuations contribute significantly to the
observed spectral dynamics. Thus, for the present case (lysozyme),
--N$_3$ serves as a highly sensitive reporter of local solvation
changes whereas --SCN appears to be better suited to provide a
connection between protein motions and recurring vibrational
environments.\\

\noindent
Overall, the results demonstrate that the vibrational probes report on
distinct aspects of the local protein environment. The --SCN probe
opens the possibility to relate local hydration, and low-frequency
protein motions, with recurrence events associated with reduced water
occupancy. The --N$_3$ probe remains strongly influenced by rapid
solvent and electrostatic fluctuations, resulting in faster FFCF
decay. Nevertheless, distinct oscillatory patterns are observed at
several labeling sites, indicating that --N$_3$ can also capture
signatures of slower, site-specific protein motions despite the
dominant contribution of rapid environmental fluctuations. In contrast
to that, the --SNO probe, exhibits site-dependent changes in its
vibrational response, and includes both oscillatory frequency
fluctuations and substantial shifts in the frequency. These shifts are
apparent in the 1D-IR lineshapes as well as in the calculated 1D-IR
spectra obtained from the probe dipole moments, where --SNO samples a
broader frequency range than --SCN and --N$_3$. This differs for
--N$_3$ and --SCN, which are in a spectroscopically ``empty'' region
of the protein-IR spectrum.\\

\noindent
Taken together, the complementary sensitivities of --SCN, --N$_3$, and
--SNO labels of alanine residues demonstrate that combining multiple
vibrational probes provides a more comprehensive view of site-specific
protein dynamics, enabling distinct contributions from hydration,
electrostatic interactions, structural packing, and conformational
fluctuations are reflected in the low-frequency vibrational modes
across the protein.\\

\section{Declaration of Interests}
The authors declare no competing interests.\\

\section{Data Availability} 
The data that support the findings of this study are available from
the corresponding author upon reasonable request.\\

\section{Acknowledgment}
Financial support from the Swiss National Science Foundation through
grants $200020\_219779$ (MM), $200021\_215088$ (MM), the University of
Basel (MM) is gratefully acknowledged.

\clearpage

\renewcommand{\thepage}{S\arabic{page}}
\renewcommand{\thetable}{S\arabic{table}}
\renewcommand{\thefigure}{S\arabic{figure}}
\renewcommand{\theequation}{S\arabic{equation}}
\renewcommand{\thesection}{S\arabic{section}} 
\setcounter{figure}{0}  
\setcounter{section}{0}  
\setcounter{table}{0}

\section*{Supporting Material}

\begin{table}[!htbp]
\centering
\caption{Atomic charges from the fMDCM fit\cite{MM.fmdcm:2022} for
  CH$_3$SCN, along with Lennard-Jones (LJ) parameters $\epsilon$
  (kcal/mol) and $r_{min}/2$ (\AA\/) obtained using the model. For the
  methyl group, LJ parameters were taken from the CGenFF force
  field.\cite{cgenff:2012}}
\begin{tabular}{c*{4}{r}*{2}{r}}
\toprule
\multicolumn{1}{c}{Atom types} & \multicolumn{4}{c}{Charges} & \multicolumn{2}{c}{LJ} \\
\cmidrule(lr){2-5} \cmidrule(lr){6-7}
 & Q1 & Q2 & Q3 & Q4 & $\epsilon$ & $r_{min}/2$ \\
\midrule
S & 0.997 & -0.890 & -0.890 & 0.488 & -0.13 & 1.50 \\
C & -0.305 & 0.844 & -- & -- & -0.14 & 2.20 \\
N & -0.439 & -0.864 & 0.864 & -- & -0.02 & 2.22 \\
C2 & 0.010  & -0.010 & -- & -- & -- & -- \\
H1 & 0.064 & -- & -- & -- & -- & -- \\
H2 & 0.064 & -- & -- & -- & -- & -- \\
H3 & 0.064 & -- & -- & -- & -- & -- \\

\bottomrule
\end{tabular}
\label{sitab:table1}
\end{table}

\begin{table}[!htbp]
\centering
\caption{Atomic charges from the fMDCM fit for CH$_3$SNO, along with
  Lennard-Jones (LJ) parameters $\epsilon$ (kcal/mol) and $r_{min}/2$
  (\AA\/) obtained using the model. For the methyl group,
  LJ-parameters were taken from the CGenFF force
  field.\cite{cgenff:2012}}
\begin{tabular}{c*{4}{r}*{2}{r}}
\toprule
\multicolumn{1}{c}{Atom types} & \multicolumn{4}{c}{Charges} & \multicolumn{2}{c}{LJ} \\
\cmidrule(lr){2-5} \cmidrule(lr){6-7}
 & Q1 & Q2 & Q3 & Q4 & $\epsilon$ & $r_{min}/2$ \\
\midrule
S & 1.000 & -0.636 & -0.636 & 0.070 & 0.40 & 2.20 \\
N & 0.463 & 0.963 & -0.236 & -- & -0.04 & 2.20 \\
O & -0.991 & -0.072 & -0.320 & -- & -0.01 & 1.99 \\
C2 & -0.761 & 0.692 & -- & -- & -- & -- \\
H1 & 0.059 & -- & -- & -- & -- & -- \\
H2 & 0.059 & -- & -- & -- & -- & -- \\
H3 & 0.059 & --& -- & -- & -- & -- \\

\bottomrule
\end{tabular}
\label{sitab:table2}
\end{table}

\begin{table}[!htbp]
\centering
\caption{Atomic charges from the fMDCM fit for CH$_3$N$_3$, along with
  Lennard-Jones (LJ) parameters $\epsilon$ (kcal/mol) and $r_{min}/2$
  (\AA\/) obtained using the model. For the methyl group,
  LJ-parameters were taken from the CGenFF force
  field.\cite{cgenff:2012}}
\begin{tabular}{c*{4}{r}*{2}{r}}
\toprule
\multicolumn{1}{c}{Atom types} & \multicolumn{4}{c}{Charges} & \multicolumn{2}{c}{LJ} \\
\cmidrule(lr){2-5} \cmidrule(lr){6-7}
 & Q1 & Q2 & Q3 & Q4 & $\epsilon$ & $r_{min}/2$ \\
\midrule
N1 & -0.931 &  0.448 & -1.000 & -- & -0.01 & 2.00 \\
N2 & 0.992 & 0.796 & -- & -- & -0.01 & 2.33 \\
N3 & -0.887 & -0.658 & 0.910 & -- & -0.01 & 2.19 \\
C2 & 0.505 & -0.301 & -- & -- & -- & -- \\
H1 & 0.042 & -- & -- & -- & -- & -- \\
H2 & 0.042 & -- &-- & -- & -- & -- \\
H3 & 0.042 & -- & -- & -- & -- & -- \\

\bottomrule
\end{tabular}
\label{sitab:table3}
\end{table}

\begin{table}[!htbp]
\centering
\begin{tabular}{ccc}
\toprule
Probe & Unfitted & Fitted \\
\midrule
--SCN & 1.32 & 1.31 \\
--SNO &  2.06 & 0.82 \\
--N$_3$ & 0.72 & 0.51 \\
\bottomrule
\end{tabular}
\caption{RMSE values (kcal/mol) of the fMDCM electrostatic models
  without and with fitted LJ-parameters starting from CGenFF
  parameters\cite{cgenff:2012} using one water molecule around the
  probe. Rows show different probes.}
\label{sitab:table4}
\end{table}

 \begin{figure}
     \centering \includegraphics[width=1\linewidth]{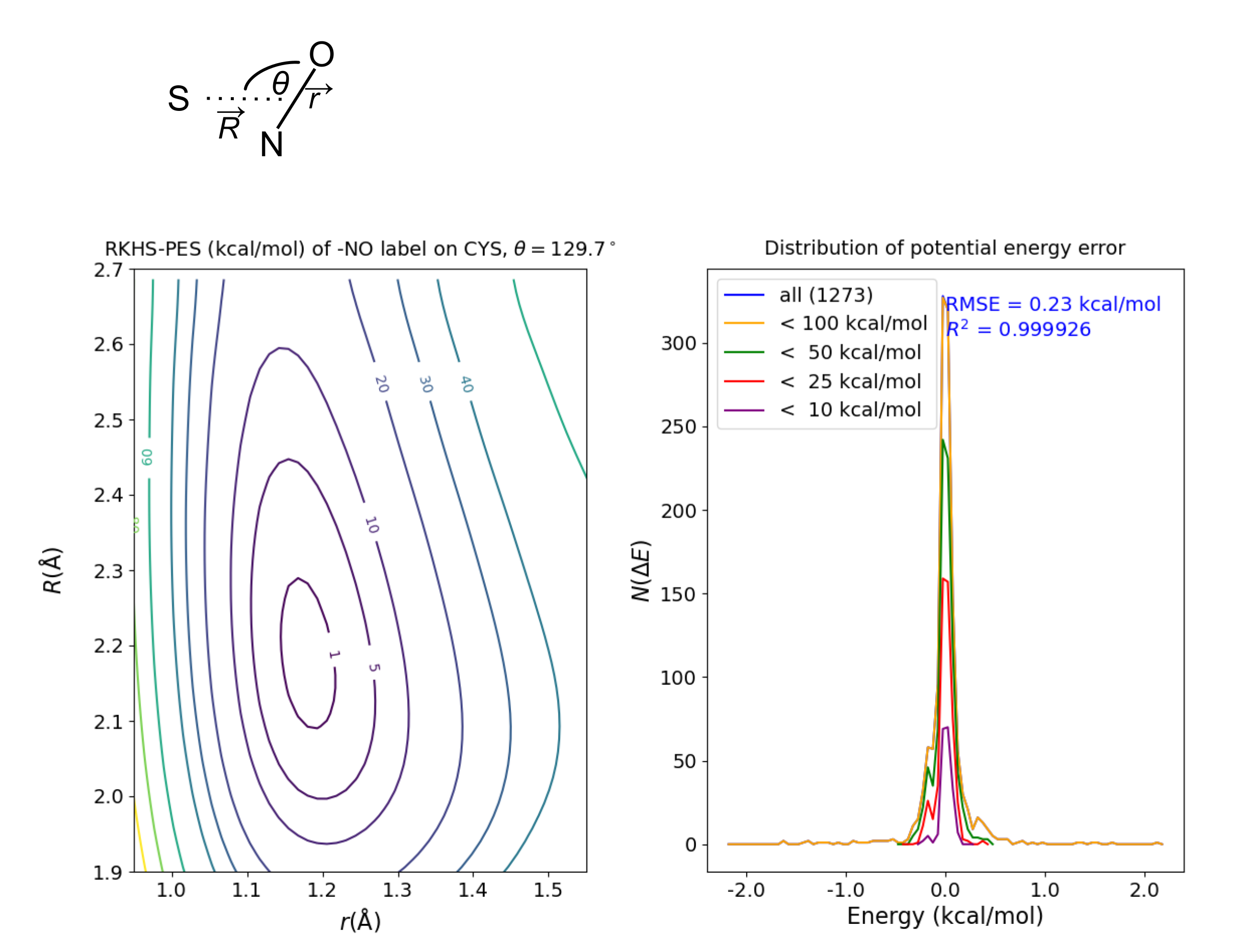}
     \caption{Left: Contour representation of the RKHS potential
       energy surface (PES) for the --SNO label based on
       PNO-LCCSD(T)-F12 $ab initio$ points in Jacobi coordinates $(R,
       r, \theta)$ at $\theta = 129.7^\circ$. A schematic illustration
       of the --SNO Jacobi coordinates and their connection to the
       alanine residue is shown in the left corner. Right: The
       distribution of potential energy errors (in kcal/mol), yielding
       an RMSE of 0.23 kcal/mol.}
     \label{sifig:fig1}
 \end{figure}

\begin{figure}
  \centering
  \includegraphics[width=0.8\linewidth]{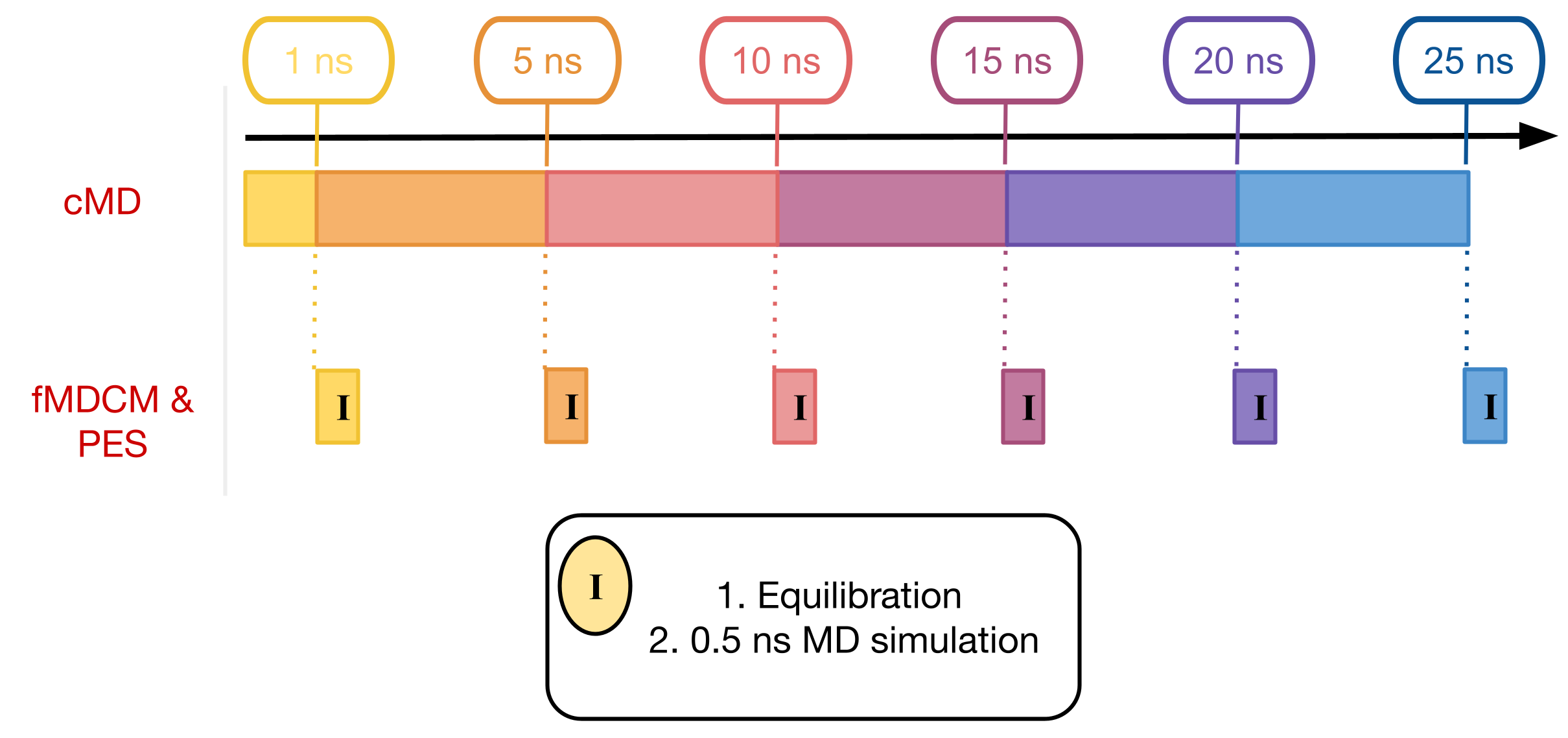}
    \caption{Workflow of 25 ns production MD simulations,
      re-equilibration from selected restart points, and subsequent
      0.5 ns simulations with fMDCM and PES models applied to labeled
      residues.}
    \label{sifig:fig2}
\end{figure}

\begin{figure}
    \centering
    \includegraphics[width=1\linewidth]{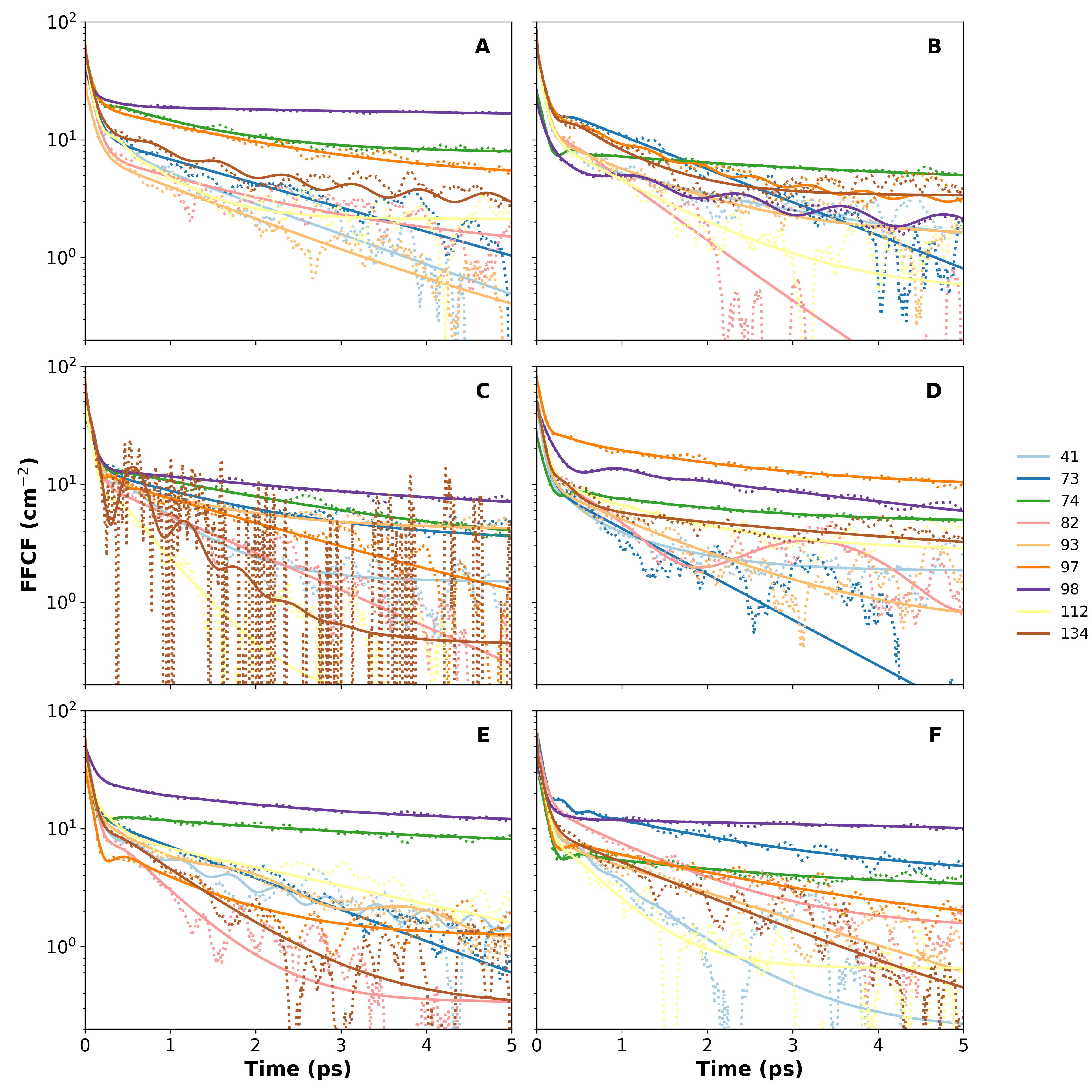}
    \caption{9 FFCFs obtained using INM analysis for the --SCN
      probe. Dashed lines are the raw data, and Solid lines show the
      fits to the Eq. \ref{eq:ffcf.fit}. The y-axis is
      logarithmic. Panels A to F correspond to simulation windows 0.5
      ns wide and the [RKHS,fMDCM]-simulations were started from
      structures after 1 (A), 5 (B), 10 (C), 15 (D), 20 (E), and 25
      (F) ns, respectively, of cMD simulations.}
    \label{sifig:fig3}
\end{figure}

\begin{longtable}{c|cccccc}
\caption{Decay times, $\tau_1$ and $\tau_2$ (in ps, upper and lower
  rows, respectively), obtained from fitting the FFCF to
  \ref{eq:ffcf.fit} for the INM frequencies of all AlaQQSCN
  residues in lysozyme. Columns correspond to the
  [RKHS,fMDCM]-simulation windows of 1, 5, 15, and 20 ns,
  respectively. The analysis was performed using a 0.5 ns-wide data
  window.}
\label{sitab:table5}\\

\toprule
\multirow{2}{*}{Residue} 
& \multicolumn{6}{c}{Time window (ns)} \\
\cmidrule(l){2-7}
& 1 & 5 & 10 & 15 & 20 & 25 \\
\midrule
\endfirsthead

\toprule
\multirow{2}{*}{Residue} 
& \multicolumn{6}{c}{Time window (ns)} \\
\cmidrule(l){2-7}
& 1 & 5 & 10 & 15 & 20 & 25 \\
\midrule
\endhead

\multirow{2}{*}{41}
& 0.2& 0.2& 0.01& 0.2& 0.08& 0.07\\
& 1.70& 1.97& 0.68& 0.91& 1.29& 0.81\\

\multirow{2}{*}{42}
& 0.17& 0.19& 0.2& 0.2& 0.2& 0.03\\
& 3.53& 5.03& 4.62& 8.04& 4.62& 2.32\\

\multirow{2}{*}{49}
& 0.09& 0.09& 0.07& 0.07& 0.2& 0.2\\
& 2.11& 2.61& 1.04& 1.62& 2.29& 2.80\\

\multirow{2}{*}{63}
& 0.04& 0.08& 0.2& 0.08& 0.06& 0.2\\
& 2.65& 3.29& 4.99& 2.68& 2.57& 6.16\\

\multirow{2}{*}{73}
& 0.09& 0.01& 0.08& 0.07& 0.07& 0.07\\
& 2.13& 1.54& 1.58& 1.12& 1.62& 1.95\\

\multirow{2}{*}{74}
& 0.01& 0.09& 0& 0.07& 0& 0.08\\
& 1.15& 3.60& 2.35& 1.76& 3.62& 2.43\\

\multirow{2}{*}{82}
& 0.08& 0.01& 0.07& 0.07& 0.01& 0.07\\
& 1.76& 0.83& 1.38& 1.35& 0.61& 1.12\\

\multirow{2}{*}{93}
& 0.09& 0.07& 0.08& 0.07& 0.07& 0.10\\
& 1.55& 1.17& 1.22& 1.30& 1.42& 1.94\\

\multirow{2}{*}{97}
& 0.08& 0.07& 0.07& 0.18& 0& 0.11\\
& 1.83& 1.17& 2.01& 1.95& 0.97& 2.22\\

\multirow{2}{*}{98}
& 0.2& 0.10& 0.10& 0.12& 0.2& 0.11\\
& 10.74& 1.79& 3.37& 4.73& 2.33& 26.80\\

\multirow{2}{*}{99}
& 0.13& 0& 0& 0.10& 0& 0\\
& 3.90& 1.22& 1.01& 7.95& 0.75& 20.59\\

\multirow{2}{*}{112}
& 0.06& 0.07& 0.01& 0.08& 0.18& 0.01\\
& 0.59& 1.03& 0.51& 1.26& 2.78& 0.54\\

\multirow{2}{*}{129}
& 0.09& 0.07& 0.01& 0.10& 0.09& 0.08\\
& 2.45& 1.99& 1.54& 5.40& 2.96& 2.79\\

\multirow{2}{*}{130}
& 0& 0.09& 0& 0.2& 0& 0.2\\
& 1.45& 11.65& 1.85& 3.54& 0.69& 2.46\\

\multirow{2}{*}{134}
& 0.07& 0.01& 0.02& 0.2& 0.01& 0.08\\
& 1.22& 0.72& 0.63& 3.67& 0.85& 1.46\\

\multirow{2}{*}{146}
& 0.12& 0& 0.10& 0.09& 0& 0\\
& 3.48& 2.73& 5.87& 2.24& 1.80& 1.36\\

\multirow{2}{*}{160}
& 0.2& 0.19& 0.07& 0.02& 0.08& 0.09\\
& 10.18& 2.68& 2.91& 1.48& 3.72& 4.04\\

\bottomrule
\end{longtable}

\begin{longtable}{c|cccccc}
\caption{Decay times, $\tau_1$ and $\tau_2$ (in ps, upper and lower
  rows, respectively), obtained from fitting the FFCF to
  \ref{eq:ffcf.fit} for the INM frequencies of all AlaQQSNO
  residues in lysozyme. Columns correspond to the
  [RKHS,fMDCM]-simulation windows of 1, 5, 15, and 20 ns,
  respectively. The analysis was performed using a 0.5 ns-wide data
  window.}
\label{sitab:table6}\\

\toprule
\multirow{2}{*}{Residue} 
& \multicolumn{6}{c}{Time window (ns)} \\
\cmidrule(l){2-7}
& 1 & 5 & 10 & 15 & 20 & 25 \\
\midrule
\endfirsthead

\toprule
\multirow{2}{*}{Residue} 
& \multicolumn{6}{c}{Time window (ns)} \\
\cmidrule(l){2-7}
& 1 & 5 & 10 & 15 & 20 & 25 \\
\midrule
\endhead

\multirow{2}{*}{41}
& 0.17& 0.03& 0.03& 0.03& 0.04& 0.03\\
& 3.27& 1.14& 0.92& 0.79& 1.41& 1.44\\

\multirow{2}{*}{42}
& 0.2& 0.16& 0.2& 0.03& 0.10& 0.03\\
& 1.90& 2.03& 3.16& 1.21& 1.33& 0.88\\

\multirow{2}{*}{49}
& 0.06& 0.03& 0.03& 0.02& 0.08& 0.03\\
& 3.06& 1.15& 1.00& 0.76& 1.74& 1.25\\

\multirow{2}{*}{63}
& 0.03& 0.08& 0.04& 0.05& 0.06& 0.05\\
& 10.64& 3.22& 2.34& 1.44& 2.38& 1.83\\

\multirow{2}{*}{73}
& 0.2& 0.03& 0.2& 0.2& 0.02& 0.03\\
& 5.14& 1.82& 3.07& 8.74& 1.38& 2.23\\

\multirow{2}{*}{74}
& 0.08& 0.14& 0.03& 0.11& 0.10& 0.11\\
& 2.44& 3.58& 1.09& 1.98& 1.31& 1.54\\

\multirow{2}{*}{82}
& 0.2& 0.2& 0.03& 0.03& 0.03& 0.06\\
& 3.07& 1.84& 1.76& 0.72& 2.11& 1.32\\

\multirow{2}{*}{93}
& 0.2& 0.2& 0.2& 0.2& 0.08& 0.13\\
& 1.11& 2.31& 5.39& 3.40& 1.04& 0.87\\

\multirow{2}{*}{97}
& 0.02& 0.03& 0.02& 0.02& 0.04& 0.2\\
& 4.05& 1.30& 3.87& 3.45& 4.03& 3.54\\

\multirow{2}{*}{98}
& 0.06& 0.09& 0.08& 0.07& 0.15& 0.07\\
& 3.29& 2.15& 3.19& 5.66& 11.79& 1.13\\

\multirow{2}{*}{99}
& 0.03& 0.06& 0.04& 0.05& 0.04& 0.03\\
& 2.55& 3.50& 16.69& 3.76& 7.87& 3.85\\

\multirow{2}{*}{112}
& 0.03& 0.2& 0.2& 0.2& 0.03& 0.03\\
& 1.83& 7.06& 2.11& 2.91& 0.93& 0.88\\

\multirow{2}{*}{129}
& 0.09& 0.09& 0.15& 0.08& 0.03& 0.03\\
& 2.69& 2.00& 2.27& 0.94& 1.13& 0.97\\

\multirow{2}{*}{130}
& 0.17& 0.01& 0.04& 0.03& 0.17& 0.03\\
& 2.65& 1.35& 1.85& 2.25& 1.90& 1.03\\

\multirow{2}{*}{134}
& 0.09& 0.2& 0.16& 0.03& 0.03& 0.03\\
& 1.53& 2.76& 1.84& 1.53& 0.92& 1.28\\

\multirow{2}{*}{146}
& 0.12& 0.03& 0.03& 0.09& 0.07& 0.03\\
& 3.99& 2.90& 3.95& 5.18& 4.39& 1.43\\

\multirow{2}{*}{160}
& 0.05& 0.03& 0.08& 0.03& 0.03& 0.02\\
& 3.11& 1.61& 1.62& 1.07& 0.69& 1.27\\

\bottomrule
\end{longtable}

\begin{longtable}{c|cccccc}
\caption{Decay times, $\tau_1$ and $\tau_2$ (in ps, upper and lower
  rows, respectively), obtained from fitting the FFCF to
  \ref{eq:ffcf.fit} for the INM frequencies of all AlaQQN$_3$
  residues in lysozyme. Columns correspond to the
  [RKHS,fMDCM]-simulation windows of 1, 5, 15, and 20 ns,
  respectively. The analysis was performed using a 0.5 ns-wide data
  window.}
\label{sitab:table7}\\

\toprule
\multirow{2}{*}{Residue} 
& \multicolumn{6}{c}{Time window (ns)} \\
\cmidrule(l){2-7}
& 1 & 5 & 10 & 15 & 20 & 25 \\
\midrule
\endfirsthead

\toprule
\multirow{2}{*}{Residue} 
& \multicolumn{6}{c}{Time window (ns)} \\
\cmidrule(l){2-7}
& 1 & 5 & 10 & 15 & 20 & 25 \\
\midrule
\endhead

\multirow{2}{*}{41}
& 0.2& 0.01& 0.2& 0.2& 0.08& 0.2\\
& 2.18& 1.60& 2.04& 4.60& 1.34& 1.38\\

\multirow{2}{*}{42}
& 0.05& 0.08& 0.08& 0.08& 0.09& 0.10\\
& 0.60& 1.93& 1.62& 1.60& 3.24& 2.56\\

\multirow{2}{*}{49}
& 0.2& 0.07& 0.2& 0.2& 0.06& 0.05\\
& 4.51& 1.27& 3.77& 5.61& 10.17& 5.30\\

\multirow{2}{*}{63}
& 0.07& 0.08& 0.08& 0.08& 0.01& 0.08\\
& 3.38& 2.01& 2.84& 2.74& 2.30& 8.71\\

\multirow{2}{*}{73}
& 0.2& 0.2& 0.07& 0.07& 0.08& 0.04\\
& 1.33& 2.52& 2.93& 1.57& 1.82& 0.76\\

\multirow{2}{*}{74}
& 0.08& 0.09& 0.07& 0.07& 0.06& 0.06\\
& 2.23& 3.12& 1.48& 1.92& 1.23& 4.27\\

\multirow{2}{*}{82}
& 0.08& 0.18& 0.2& 0.07& 0.2& 0.08\\
& 1.12& 2.73& 1.63& 1.02& 0.76& 2.01\\

\multirow{2}{*}{93}
& 0.2& 0.2& 0.01& 0.2& 0.01& 0.06\\
& 3.87& 6.11& 1.38& 3.21& 1.67& 2.55\\

\multirow{2}{*}{97}
& 0.07& 0.2& 0.06& 0.01& 0.06& 0.10\\
& 2.75& 4.33& 3.00& 4.01& 5.29& 3.09\\

\multirow{2}{*}{98}
& 0.08& 0.09& 0.2& 0& 0& 0.05\\
& 2.90& 4.80& 3.77& 3.25& 1.73& 2.56\\

\multirow{2}{*}{99}
& 0.08& 0.2& 0.2& 0.06& 0.10& 0.08\\
& 2.68& 2.50& 5.05& 2.05& 6.04& 2.54\\

\multirow{2}{*}{112}
& 0.01& 0.2& 0.09& 0.2& 0.09& 0.08\\
& 1.09& 10.11& 1.31& 1.11& 2.24& 1.49\\

\multirow{2}{*}{129}
& 0.06& 0.06& 0.08& 0.04& 0.06& 0.04\\
& 2.69& 4.57& 3.33& 2.83& 1.93& 1.97\\

\multirow{2}{*}{130}
& 0.07& 0.2& 0.07& 0.2& 0.2& 0.08\\
& 1.80& 3.37& 1.88& 4.23& 2.68& 3.98\\

\multirow{2}{*}{134}
& 0.2& 0.2& 0.06& 0.07& 0.01& 0.2\\
& 6.39& 1.99& 2.87& 2.57& 2.73& 3.74\\

\multirow{2}{*}{146}
& 0.13& 0.04& 0.11& 0.2& 0.13& 0.12\\
& 8.81& 0.46& 4.52& 2.40& 2.95& 4.15\\

\multirow{2}{*}{160}
& 0.2& 0.2& 0.2& 0.06& 0& 0.07\\
& 5.60& 4.34& 1.22& 2.10& 1.02& 2.70\\

\bottomrule
\end{longtable}

\begin{figure}
    \centering
    \includegraphics[width=0.8\linewidth]{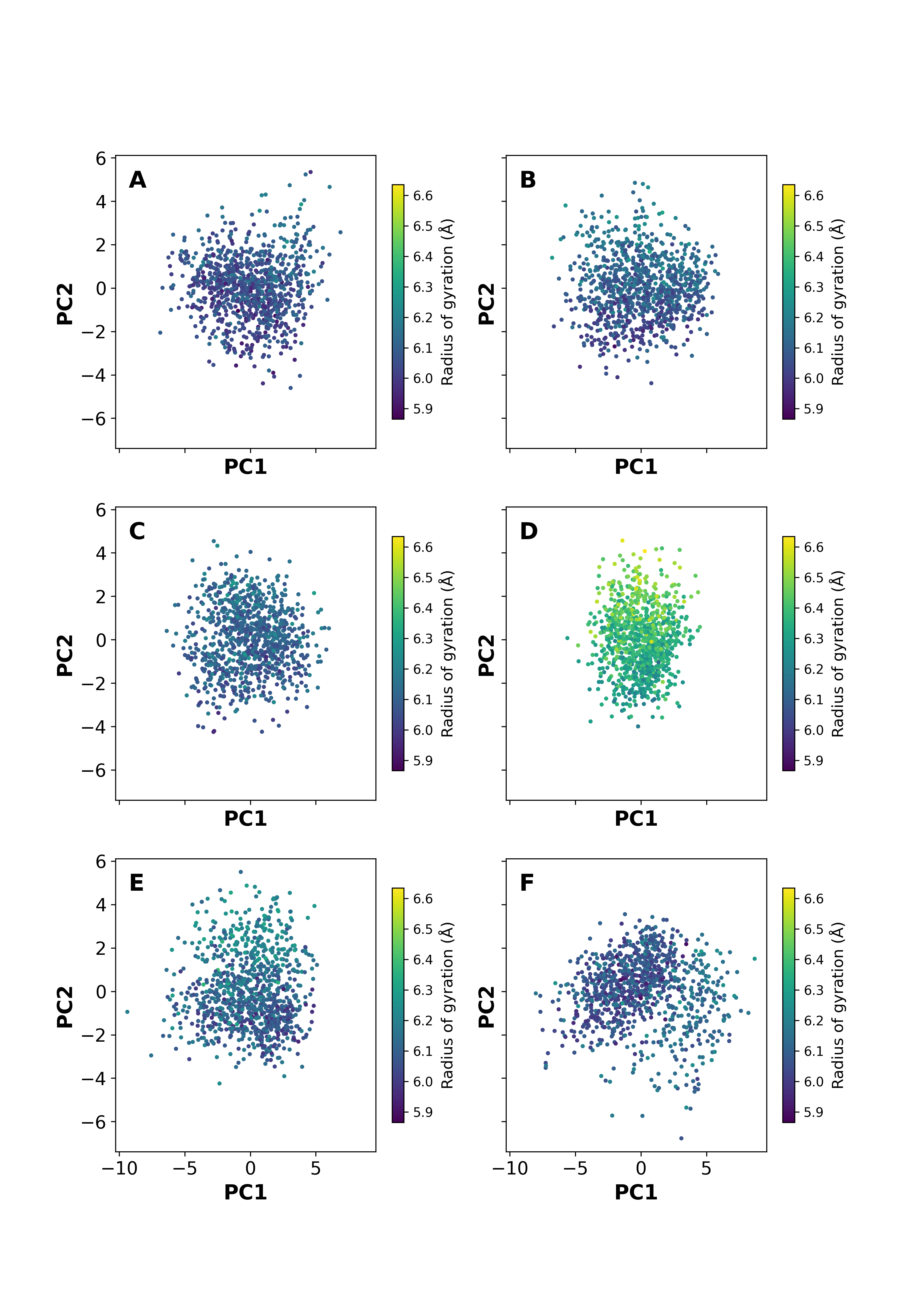}
    \caption{PCA projections of protein–probe distance descriptors for
      ALA99SCN obtained from six consecutive windows of the
      trajectory: 1 ns (A), 5 ns (B), 10 ns (C), 15 ns (D), 20 ns (E),
      and 25 ns (F). Input features for the PCA were the instantaneous
      distances between the center of mass of the SCN probe and all
      non-hydrogen protein atoms located within 7 \AA\/ of the probe
      in the backbone-aligned trajectory. Each point represents on
      frame (1000 frames per panel, 6000 frames in total), with frames
      saved every 0.5 ps. Points are colored according to the radius
      of gyration of the local protein environment used for the
      distance descriptors. The projections illustrate how
      fluctuations in the probe surroundings are associated with local
      structural expansion and compaction. Note that a PCA analysis
      for the structures for each time window were done separately,
      whereas Figure \ref{sifig:fig5} analyses all structures together
      which allows direct comparison between them.}
    \label{sifig:fig4}
\end{figure}

\begin{figure}
    \centering
    \includegraphics[width=1\linewidth]{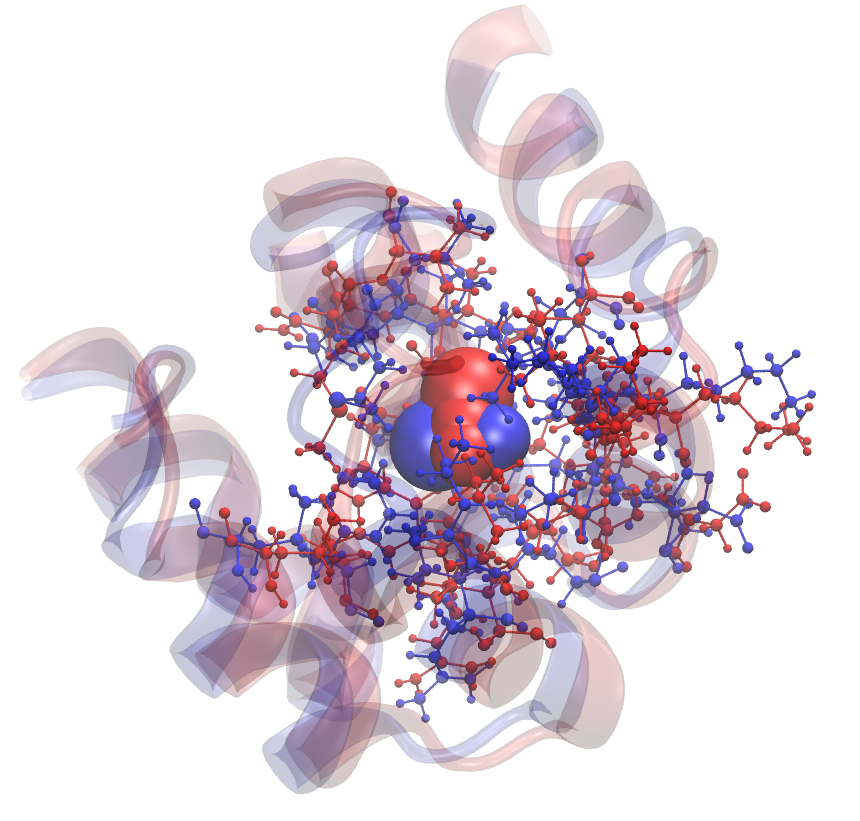}
    \caption{Representative conformational states extracted from PCA
      of protein--SCN distances for ALA99SCN. Two frames represented as
      red and blue colors with similar PC2 values but opposite PC1
      values were selected, see the stars in Figure \ref{sifig:fig4},
      RMSD-aligned using the protein backbone, and visualized in
      3D. The probe is shown in van der Waals representation, the
      protein in NewCartoon representation, and heavy atoms of
      residues surrounding the probe are displayed in CPK
      representation. These structures illustrate distinct
      probe–protein interaction states corresponding to opposite sides
      of PC1 while maintaining similar PC2 values. PC1 is primarily
      related to the orientation of the --SCN label. Protein backbone
      RMSD between the states and protein within 7 \AA\/ of the center
      of mass of --SCN probe RMSD between the states are 2.02 \AA\/ and
      1.53 \AA\/, respectively.}
    \label{sifig:fig6}
\end{figure}

\begin{figure}
    \centering
    \includegraphics[width=1\linewidth]{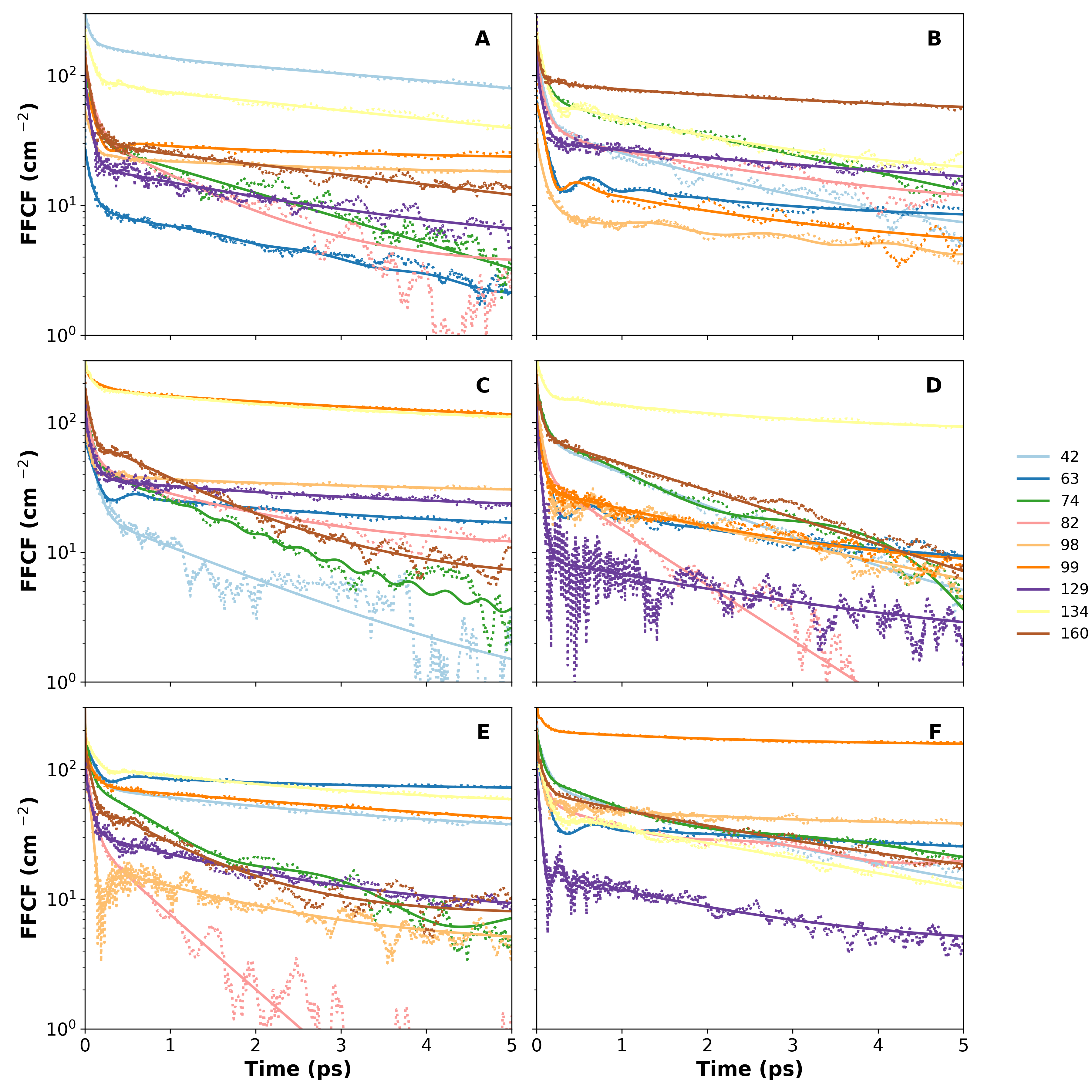}
    \caption{9 FFCFs obtained using INM analysis for the --N$_3$
      probe. Dashed lines are the raw data, and solid lines show the
      fits to the Eq. \ref{eq:ffcf.fit}. The y-axis is
      logarithmic. Panels A to F correspond to simulation windows 0.5
      ns wide and the [RKHS,fMDCM]-simulations were started from
      structures after 1 (A), 5 (B), 10 (C), 15 (D), 20 (E), and 25
      (F) ns, respectively, of cMD simulations.}
    \label{sifig:fig3-N3-1}
\end{figure}

\begin{figure}
    \centering
    \includegraphics[width=1\linewidth]{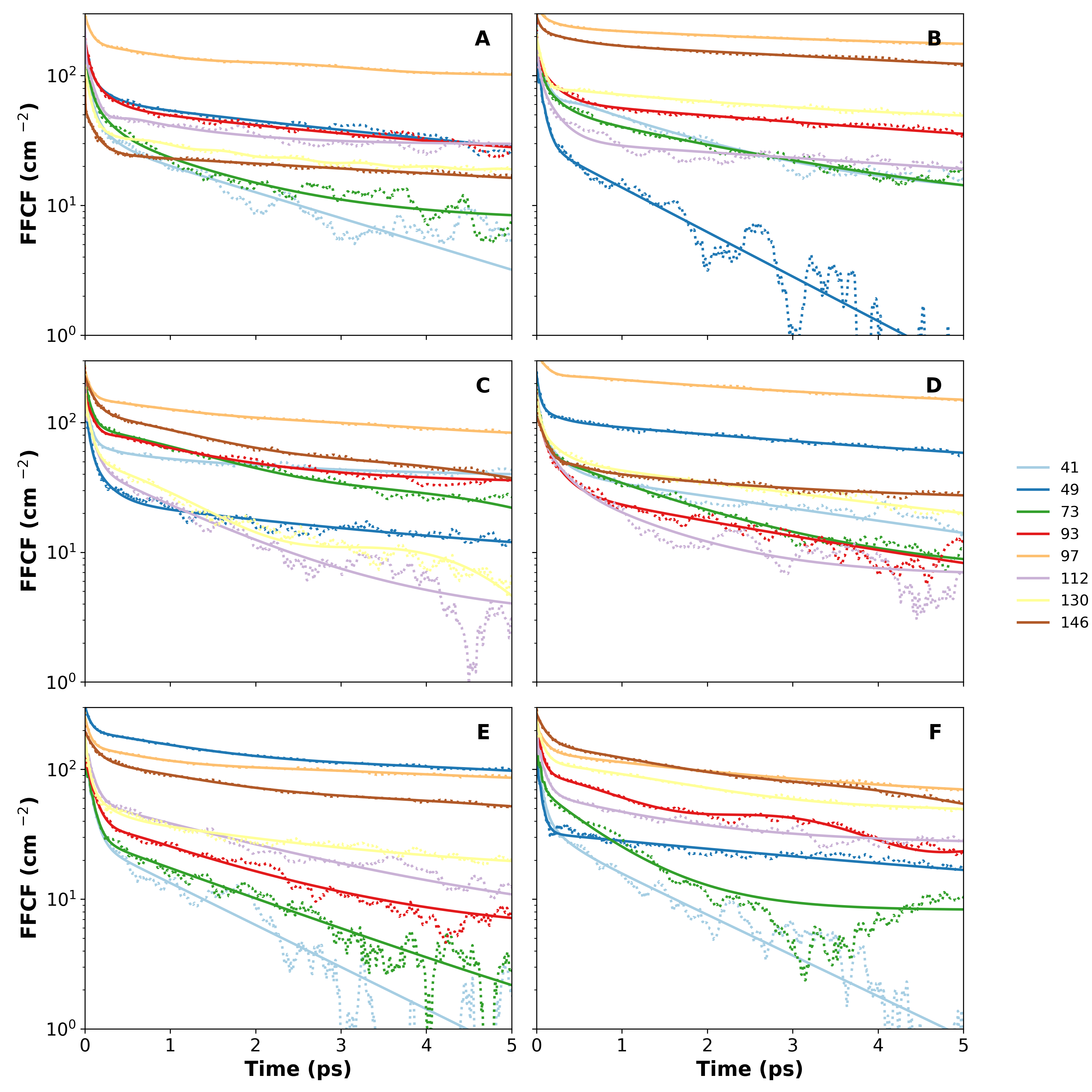}
    \caption{8 FFCFs obtained using INM analysis for the --N$_3$
      probe. Dashed lines are the raw data, and solid lines show the
      fits to the Eq. \ref{eq:ffcf.fit}. The y-axis is
      logarithmic. Panels A to F correspond to simulation windows 0.5
      ns wide and the [RKHS,fMDCM]-simulations were started from
      structures after 1 (A), 5 (B), 10 (C), 15 (D), 20 (E), and 25
      (F) ns, respectively, of cMD simulations.}
    \label{sifig:fig3-N3-2}
\end{figure}

\clearpage

\bibliography{refs}

\end{document}